\documentclass[a4paper,11pt]{article}

\usepackage{jheppub}
\usepackage{graphicx}
\usepackage{multirow}
\usepackage{bm}
\usepackage{braket}
\usepackage{color}
\usepackage{slashed}
\usepackage{diagbox}
\usepackage{booktabs}
\usepackage{array}

\title{Next-to-leading-order QCD corrections to $S$- and $P$-wave heavy quarkonium decay to $l^{+}l^{-} \gamma$ }

\author[a]{Zhi-Guo He,}
\emailAdd{zhiguo.he@buct.edu.cn}

\author[b]{Xu-Dong Huang}
\emailAdd{huangxd@cqnu.edu.cn}

\affiliation[a]{Department of Physics and Electronics, School of Mathematics and Physics, Beijing University of Chemical Technology, Beijing 100029, China}
\affiliation[b]{College of Physics and Optoelectronic Engineering, Chongqing Normal University, Chongqing 401331, P.R. China}

\abstract{
In this work, we comprehensively study the total and differential decay widths of the 
radiative Dalitz decays $H \to l^+l^-\gamma$ up to QCD next-to-leading order (NLO) 
accuracy within the framework of NRQCD factorization. Our calculation includes the 
decays of $S$-wave states ($\eta_c, \eta_b$) and the $P$-wave triplets 
($\chi_{cJ}, \chi_{bJ}$ for $J=0,1,2$) to both electron ($l=e$) and muon ($l=\mu$) 
final states. To match realistic experimental detection thresholds, systematic 
kinematic cuts are implemented on the final-state photon energy. Our analysis of the 
lepton-pair invariant mass distribution shows distinct singular behaviors across the 
multiplets originating from the lepton-pair threshold region, which is regularized by 
the lepton mass, and the soft-photon region, respectively. For the $S$-wave states, 
there is only one peak near the lepton-pair threshold region, while for the $J=0$ and 
$J=2$ $P$-wave states, the peaks show up in both regions. However, for the $J=1$ 
$P$-wave states, the peak only appears in the soft-photon region. Such features provide 
a rich venue to probe the $\gamma^{\ast}\to l^{+}l^{-}$ form factor in heavy quarkonium 
decay. Integrating over the bounded phase spaces reveals a distinct hierarchy among the 
$\chi_{QJ}$ states in the sensitivity of our theoretical predictions to the soft-photon 
energy cuts, ordered as $\chi_{Q1} >\chi_{Q2}>\chi_{Q0}$. In the
$\chi_{cJ}\to l^{+}l^{-}\gamma$ cases, as the energy cut increases from 100 $\mathrm{MeV}$, to 500 $\mathrm{MeV}$ the theoretical predictions at QCD NLO are reduced by $6\%(\chi_{c0})$,$66\%(\chi_{c1})$, and $21\%(\chi_{c2})$ for $l=e$, and by $17\%(\chi_{c0})$,$66\%(\chi_{c1})$, and $39\%(\chi_{c2})$ for $l=\mu$. Comparing our predictions with the upcoming high-precision experimental tests at BESIII will definitely deepen our understanding of the predictive power of perturbative calculations.
} 

\begin{document}

\maketitle

\flushbottom

\section{Introduction}
The hierarchy of energy scales inside heavy quarkonia $m_Qv^2\ll m_Qv\ll m_Q$, where $v$ 
is the velocity of the heavy quark in the meson rest frame, makes them a perfect 
laboratory for investigating both perturbative and non-perturbative aspects of Quantum 
Chromodynamics (QCD). The non-relativistic QCD (NRQCD) factorization 
formalism~\cite{Bodwin:1994jh} is currently the most promising approach to
describe heavy quarkonium production and decay. In NRQCD factorization, the production or 
decay rates are separated into process dependent short distance coefficients (SDCs) and 
non-perturbative long distance matrix elements (LDMEs). The validity of NRQCD 
factorization has been extensively examined from both experimental and theoretical 
perspectives; for a review see Ref.~\cite{Brambilla:2010cs}. Among the 
production and decay processes, the exclusive production or decay through 
electromagnetic interaction is of particular interest, since in such processes the 
experimental signals are relatively clean and the theoretical mechanisms are more 
transparent.

For $C=+$ states $H$, such as $S$-wave $\eta_Q$ and $P$-wave $\chi_{QJ}$, they couple 
directly with two photons. In recent years their production through 
$e^{+}e^{-}\to\gamma^{\ast} \to H+\gamma$ and $\gamma\gamma^{\ast}\to H$ has attracted 
considerable interest. The $e^{+}e^{-}\to\gamma^{\ast} \to H+\gamma$ process was 
measured by BES~\cite{BESIII:2014uzr,BESIII:2021yal} and BELLE~\cite{Belle:2018jqa} 
Collaborations around $\sqrt{s}=4$ and 10.6 GeV, respectively. Furthermore it was also 
proposed to search for the exotic $``XYZ"$ states through 
$e^{+}e^{-}\to H+\gamma$~\cite{Li:2009ki,Li:2013nna,Chao:2013cca,Yuan:2015kya} and 
for the first time the BESIII Collaboration observed the famous state $X(3872)$ in 
$e^{+}e^{-}\to X(3872)+\gamma$~\cite{BESIII:2013fnz}. Theoretically, the 
$e^{+}e^{-}\to\gamma^{\ast} \to \eta_c+\gamma$ process was first studied in 
Ref.~\cite{Shifman:1980dk} as a probe of the $\eta_c$ electromagnetic form factor, where 
the next-to-leading order QCD corrections were also included. The leading order (LO) NRQCD 
prediction of $P$-wave $\chi_{QJ}$ production via 
$e^{+}e^{-}\to\gamma^{\ast} \to \chi_{QJ}+\gamma$ was presented in 
Ref.~\cite{Chung:2008km}, and later it was found that the next-to-leading order (NLO) 
QCD~\cite{Li:2009ki,Sang:2009jc} and relativistic 
corrections~\cite{Sang:2009jc,Fan:2012dy,Xu:2014zra,Brambilla:2017kgw} can not be 
ignored. Now the theoretical calculation has arrived at QCD next-to-next-to leading order 
(NNLO) accuracy~\cite{Chen:2017pyi,Yu:2020tri,Sang:2020fql,Li:2025pbt}. When 
$\sqrt{s}\gg m_Q^2$, the large logarithm of the type $\ln s/m_Q^2$ can be resummed to all 
orders to improve the convergence of fixed order 
calculation~\cite{Jia:2008ep,Chung:2019ota}. In the photoproduction, 
$\gamma\gamma^{\ast}\to \eta_c$ was measured by the BaBar Collaboration~\cite{BaBar:2010siw}. 
The differential cross section is proportional to the form factor $F(Q^2)$ in space-like 
momentum transfer region. However, the NRQCD predictions up to QCD 
NNLO~\cite{Feng:2015uha} largely overshot the experimental data. The large discrepancy 
might be resolved by applying the Principle of Maximum Conformality to the 2-loop 
results~\cite{Wang:2018lry}, and it was shown that after combining the NNLO QCD 
corrections with relativistic corrections to all orders of $v^2$ NRQCD predictions become 
consistent with BaBar data within errors~\cite{Babiarz:2025agk}. The photon production of 
$P$-wave state $\chi_{QJ}$ has not been measured yet. Interestingly, Belle reported the 
evidence of $\gamma\gamma^{\ast}\to X(3872)$~\cite{Belle:2020ndp}, which is greatly helpful in understanding its structure~\cite{Babiarz:2023ebe}. 

Regarding the electromagnetic decay, the primary channel is $H\to \gamma\gamma$ (except 
for $\chi_{Q1}$), where the energies of the photons are fixed at $m_H/2$ in the meson 
rest frame; thus, it yields limited phenomenological information beyond the total decay 
width. The state-of-the-art NRQCD predictions for
$\Gamma(\eta_c\to \gamma\gamma)$~\cite{Feng:2017hlu} and the ratio 
$\Gamma(\chi_{c2}\to \gamma\gamma)/\Gamma(\chi_{c0}\to \gamma\gamma)$~\cite{Sang:2015uxg} 
lie significantly below the worldwide PDG data~\cite{ParticleDataGroup:2024cfk}. 
Recently, the severe tension in $\Gamma(\eta_c\to \gamma\gamma)$ was clarified by the 
latest BESIII measurements~\cite{BESIII:2026pff}, but no further results for 
$\chi_{c0,2}\to \gamma\gamma$ have been reported yet. 

To better understand the electromagnetic decay, more information can be extracted from 
the Dalitz decays of $H\to \gamma\gamma^{\ast}\to \gamma l^{+}l^{-}$, where the energy of 
the photon ranges from 0 to $m_H/2-2m_l^2/m_H$. Analogous to the production process, we 
can also probe the form factor in the time-like momentum transfer region
$4m_l^2<Q^2<m_H^2$~\cite{DiSalvo:2000ec}. Moreover, such processes can also serve to 
test the NRQCD factorization formalism. Particularly in the $P$-wave case, perturbative 
calculations at LO will encounter infrared (IR) divergences, which is very similar to the 
decay of $\chi_{bJ}$ into charmed hadrons~\cite{Bodwin:2007zf}; and we recall that the IR 
divergence plaguing $\chi_{bJ}\to l^{+}l^{-}$ has been removed successfully within NRQCD 
factorization by taking into account the contribution from higher Fock states 
$Q\bar{Q}(^3S_1^{[1]})$~\cite{Yang:2012gk,Kivel:2015iea,Jia:2024dzm}. Additionally, the 
Dalitz decays are also the main background of the rare decay of $H\to l^{+}l^{-}$. All 
these above considerations motivate us to investigate the Dalitz decays of $\eta_Q$ and $\chi_{QJ}$ to 
$\gamma l^{+}l^{-}$. 

Theoretically, only the $\eta_Q\to \gamma l^{+}l^{-}$ was calculated at QCD 
LO~\cite{Jia:2009ip}. In this work, we will systematically study the Dalitz decays of 
$\eta_Q$ and $\chi_{QJ}$ to $\gamma l^{+}l^{-}$. It is well known that higher order QCD 
corrections may be important in heavy quarkonium production and decay, hence we will 
compute the NLO QCD corrections as well. To date, no experimental measurements have been
carried out. It is worth noting that the nature of $X(3872)$ remains poorly 
understood. In some models, it is believed to possess a considerable $\chi_{c1}(2P)$ 
component~\cite{Li:2009ad,Wang:2015rcz,Tan:2019qwe,Man:2024mvl}. Should our predictions align with future experimental data for 
$\chi_{c1}\to \gamma l^{+}l^{-}$, the corresponding study of 
$X(3872)\to \gamma l^{+}l^{-}$ could greatly aid in understanding the nature of 
$X(3872)$. 

The remainder of the paper is organized as follows. In section~\ref{I}, we describe the 
framework in our calculation and provide the relevant formulas that we use. In section~\ref{II}, we will dedicate to present the numerical results and discussions. In the end, a summary will be given in section~\ref{III}.

\section{Calculation technology} \label{I}

\subsection{General formalism}

Within the framework of NRQCD factorization, the Dalitz decay of $H=\eta_Q$, and $\chi_{QJ}$ into $l^{+}l^{-} \gamma$ can be expressed as:
\begin{align}
d\Gamma_{H \to l^{+}+l^{-}+\gamma}=d{\hat \Gamma}_{(Q\bar{Q})[n] \to l^{+}+l^{-}+\gamma}\langle H| {\cal O}(n)|H\rangle,
\label{nrqcdfact}
\end{align}
where $d\hat{\Gamma}$ encodes the perturbatively calculable SDC and 
$\langle H| {\cal O}(n)|H\rangle$ is the corresponding universal LDME. For the 
electromagnetic decay, at LO in $v^2$ only the color-singlet sector contributes, 
requiring explicit consideration of only $n=$ $^1S_0$ and $^3P_J$ ($J=0,1,2$) 
spectroscopic configurations, and the relevant four-fermion operators are defined as
~\cite{Bodwin:1994jh}\footnote{Here we include an additional normalization factor of \(1/(2N_c)\) compared to the original definitions in Ref~\cite{Bodwin:1994jh}.}
\begin{subequations}
	\begin{align}
		&{\cal O}(^1S_0)=\frac{1}{2N_c} \psi^\dagger\chi |0\rangle\langle0|\chi^\dagger \psi,&\\
		&{\cal O}(^3P_0)=\frac{1}{6N_c}\psi^\dagger \left(-\frac{i}{2}\overleftrightarrow{{\bf D}}\cdot {\bm\sigma}\right)\chi |0\rangle\langle0|\chi^\dagger\left(-\frac{i}{2}\overleftrightarrow{{\bf D}}\cdot {\bm\sigma}\right) \psi,&\\
		&{\cal O}(^3P_1)=\frac{1}{4N_c}\psi^\dagger \left(-\frac{i}{2}\overleftrightarrow{{\bf D}}\times {\bm\sigma}\right) \chi |0\rangle\langle0|\chi^\dagger \left(-\frac{i}{2}\overleftrightarrow{{\bf D}}\times {\bm\sigma}\right) \psi,&\\
		&{\cal O}(^3P_2)=\frac{1}{2N_c}\psi^\dagger \left(-{i\over 2}\overleftrightarrow{\bf{D}}^{(i}\bm\sigma^{j)} \right)\chi|0\rangle\langle0|\chi^\dagger \left( -{i\over 2}\overleftrightarrow{\bf{D}}^{(i}\bm\sigma^{j)}\right)\psi,&
	\end{align}
\end{subequations}
where $\overleftrightarrow{{\bf D}}=\overrightarrow{{\bf D}}-\overleftarrow{{\bf D}}$, 
$\overleftrightarrow{\bf{D}}^{(i}\bm\sigma^{j)}=(\overleftrightarrow{\bf{D}}^{i}\bm\sigma^{j}+\overleftrightarrow{\bf{D}}^{j}\bm\sigma^{i})/2-\frac{\delta^{ij}}{3}\overleftrightarrow{\bf{D}}\cdot\bm\sigma$, and $\psi(\chi^\dagger)$ represents the Pauli-spinor field 
that annihilates a heavy (antiquark) quark. The LDMEs can be determined via lattice 
simulations, phenomenological fits to experimental data, or by relating them to the wave 
function at the origin in potential model calculations for color-singlet states. At $v^2$ LO their relations with the wave functions are: 
\begin{eqnarray}\label{LDMEs}
\langle H| {\cal O}(^1S_0^{[1]})|H\rangle=\frac{|R_S(0)|^2}{4\pi}, 
\langle H| {\cal O}(^3P_J^{[1]})|H\rangle=\frac{3|R_P^{\prime}(0)|^2}{4\pi},
\end{eqnarray}

The SDCs can be determined by matching 
between perturbative QCD and NRQCD calculation for free $Q\bar{Q}$ decay in Fock state 
$n$, yielding
\begin{eqnarray}\label{width}
	d\Gamma_{(Q\bar{Q})[n] \to l^{+}+l^{-}+\gamma}|_{\mathrm{per\; QCD}}=d{\hat \Gamma}_{(Q\bar{Q})[n] \to l^{+}+l^{-}+\gamma}\langle Q\bar{Q}|{\cal O}(n)|Q\bar{Q}\rangle. \label{sdcs}
\end{eqnarray}
The left-hand of Eq.(\ref{sdcs}) can be computed straightforwardly with the help of 
spinor projection method~\cite{Petrelli:1997ge}, in which the product of Dirac spinors 
$u(p_Q)\bar{v}(p_{\bar{Q}})$ is projected onto the configuration of spin-singlet ($S=0$) 
or spin-triplet ($S=1$) with vector $\epsilon_S^{\mu}$ in a Lorentz covariant 
form~\cite{Petrelli:1997ge}
\begin{subequations}
\begin{eqnarray}
\Pi_0&=&\sum_{s_1,s_2}u(p_Q)\bar{v}(p_{\bar{Q}})\langle \frac{1}{2},s_1;\frac{1}{2},s_2|0,0\rangle=
\frac{1}{2\sqrt{2}m_Q}(
\slashed{p}_{\bar Q}+m_Q)\gamma^5(\slashed{p}_Q-m_Q),\\
\Pi_1^{\mu}&=&\sum_{s_1,s_2}u(p_Q)\bar{v}(p_{\bar{Q}})\langle \frac{1}{2},s_1;\frac{1}{2},s_2|1,S_z\rangle
=\frac{1}{2\sqrt{2}m_Q}(\slashed{p}_{\bar Q}+m_Q)\gamma^\mu(\slashed{p}_Q-m_Q).
\end{eqnarray} 
\end{subequations}
In the expressions above, the $v^2$ higher order effects in the projector are omitted. 
Similarly, the projector in $SU(3)_c$ space for CS state is
\begin{eqnarray*}
\mathcal{C}_1=\langle 3,i;\bar{3},j|1\rangle=\frac{\delta_{ij}}{\sqrt{N_c}}.
\end{eqnarray*}

To further match the amplitude on $S$- or $P$-wave state decay, it is more convenient 
to express the four-momenta of heavy quark ($Q$) and anti-quark ($\bar{Q}$) in terms of 
their total momentum $p_0$ and relative momentum $q$ as:
	\begin{align}
		p_Q=\frac{p_0}{2}+q, \;\;\;\;
		p_{\bar Q}=\frac{p_0}{2}-q,
	\end{align}
Since we are working at LO of $v^2$, we have $p_0^2=m_H^2=4m_Q^2$ in the non-relativistic 
limit. In such a way, the hard part of the full QCD calculation of $n=$$^1S_0^{[1]}$ and 
$n=$$^3P_J^{[1]}$ decay become
\begin{subequations}
\begin{eqnarray}
	\mathcal{A}_{^1S_0^{[1]}}&=&{\rm Tr}[\mathcal{C}_1\otimes\Pi_0 \mathcal{M}]\Big |_{q=0},
	\\
	\mathcal{A}_{^3P_J^{[1]}}&=&\epsilon^{(J)}_{\mu\nu}\frac{d}{dq_\nu}{\rm Tr}[\mathcal{C}_1\otimes\Pi_1^\mu \mathcal{M}]\Big |_{q=0},
\end{eqnarray}	
\end{subequations}
where $\mathcal{M}$ is the standard parton-level Feynman amplitude of 
$(Q\bar{Q}) \to l^{+}(p_1)+l^{-}(p_2)+\gamma(p_3)$ with the external quark spinors being 
truncated, and $\epsilon^{(J)}_{\mu\nu}$ is the polarization tensor of the spin triplet 
$P$-wave state $^3P_J$. 

Finally, we are in position to write down the master formulas of the SDCs for Dalitz decay through matching
\begin{eqnarray}\label{SDCs}
	d\hat{\Gamma}_{(Q\bar{Q})[n] \to l^{+}+l^{-}+\gamma}= \dfrac{1}{2m_H}\dfrac{1}{2J+1}\sum \frac{1}{m_Q}\vert \mathcal{A}_{n} \vert^2 d\Phi_3, \label{dsigma}
\end{eqnarray}
where the symbol $\sum$ is understood to sum over the polarization of both initial and 
final states, and the factor $1/m_Q$ comes from the normalization of NRQCD operators. 
$d\Phi_3$ is the three-body phase space integral. We parameterize it in the standard 
Dalitz plot form by introducing the invariant mass $m^2_{12}=(p_1+p_2)^2$ and 
$m^2_{23}=(p_2+p_3)^2$, and explicitly
\begin{eqnarray}
d\Phi_3=\frac{1}{(2\pi)^3}\frac{1}{16m_H^2}dm_{12}^2 dm_{23}^2.
\end{eqnarray}

\subsection{Calculation of the SDC up to $\mathcal{O}(\alpha_s)$}
Throughout our computation, we employ the package \texttt{FeynArts}~\cite{Hahn:2000kx} to 
generate Feynman diagrams and amplitudes, and the package 
\texttt{CalcLoop}\footnote{\texttt{CalcLoop} can be found in 
https://gitlab.com/multiloop-pku/calcloop.} 
to handle the contraction of Lorentz index, traces of Dirac and $\mathrm{SUN(3)_c}$-color 
matrices, and partial fraction. The packages \texttt{AMFlow}~\cite{Liu:2017jxz, Liu:2020kpc, Liu:2021wks, Liu:2022chg, Liu:2022mfb} and \texttt{Kira}~\cite{Klappert:2020nbg} are 
employed for integration-by-parts (IBP) reduction. Following the IBP reduction, all 
one-loop integrals are reduced to master integrals, which are numerically evaluated using 
the package \texttt{LoopTools}~\cite{Hahn:1998yk}. The final phase-space integrations 
are computed with assistance of the package \texttt{Vegas}~\cite{Lepage:1977sw}.

\begin{figure}[htbp]
\centering
\includegraphics[width=0.9\textwidth]{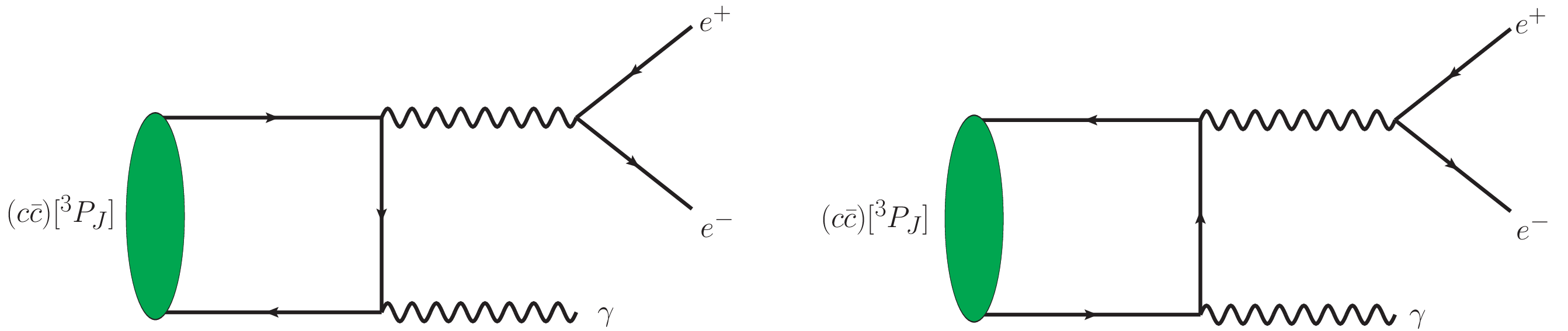}
\caption{ The LO Feynman diagrams for $(c\bar{c})[^3P_J] \to e^{+}+e^{-}+\gamma$.} \label{diaglo}
\end{figure}

\begin{figure}[htbp]
\centering
\includegraphics[width=0.9\textwidth]{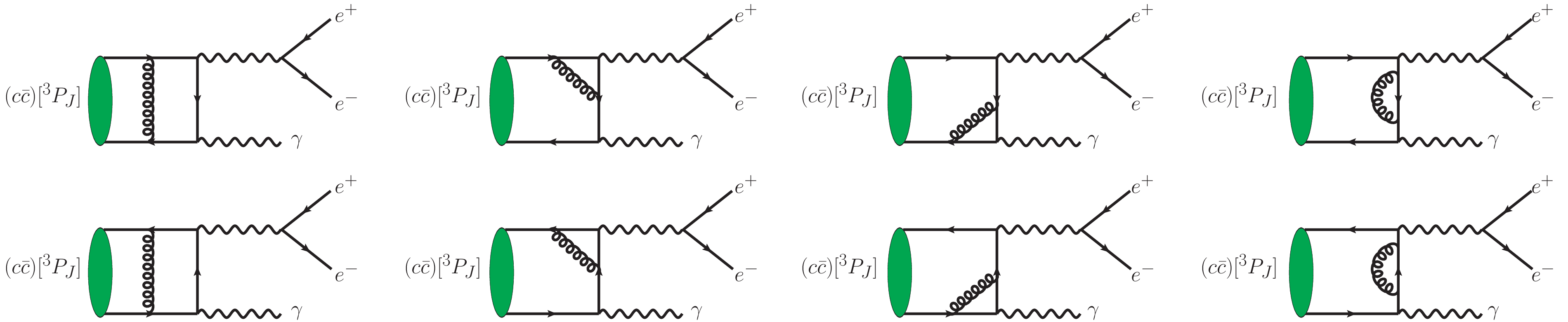}
\caption{The NLO Feynman diagrams for $(c\bar{c})[^3P_J] \to e^{+}+e^{-}+\gamma$.} \label{diagnlo}
\end{figure}

For the exclusive Dalitz decay only the one-loop diagrams contribute at QCD NLO. The LO 
and NLO Feynman diagrams are shown in Fig.\ref{diaglo} and Fig.\ref{diagnlo}, 
respectively. To systematically regulate ultraviolet (UV) and infrared (IR) divergences, 
we implement conventional dimensional regularization with spacetime dimension 
$d=4-2\epsilon$. In principle, the four-point box diagrams also contain Coulombic 
singularity, manifesting as power-law divergences in the $q\to0$ limit. This kind of 
singularity can be rigorously absorbed into the NRQCD LDME at QCD NLO~\cite{Bodwin:1994jh} 
or will not show up if we set relative momentum $q$ to zero prior to the loop integration 
in dimensional regularization. The UV divergence will be removed solely through the 
renormalization of the heavy quark mass and the field, for only these quantities are 
involved in the LO calculation. We adopt on-shell (OS) renormalization scheme for them, 
as it is the most widely used in computation of heavy quarkonium production and decay. 
The renormalization constant $Z_{2}$ for heavy quark field and $Z_{m}$ for heavy quark 
mass at one-loop level are given by
\begin{subequations}
\begin{eqnarray}
\delta Z_2^{\mathrm{OS}}&=&-C_F\frac{\alpha_s}{4\pi}N_\epsilon(\frac{1}{\epsilon_{\mathrm{UV}}}+\frac{2}{\epsilon_{\mathrm{IR}}}+4),\\
\delta Z_m^{\mathrm{OS}}&=&-3C_F\frac{\alpha_s}{4\pi}N_\epsilon(\frac{1}{\epsilon_{\mathrm{UV}}}+\frac{4}{3}),
\end{eqnarray} 
\end{subequations}
where $C_F=\frac{N_c^2-1}{2N_c}$ and $N_{\epsilon}=(4\pi\mu_R^2/m_Q^2)^{\epsilon}/\Gamma(1-\epsilon)$ with $\mu_R$ being the renormalization scale. 

The Lehmann–Symanzik–Zimmermann (LSZ) reduction formula implies that the renormalized amplitude can be written as
\begin{eqnarray}
\mathcal{A}(\alpha_s,m_Q)=Z_{2}^{\mathrm{OS}}\bigg[\mathcal{A}_{bare}^{0l}(m_{Q,bare})+\mathcal{A}_{bare}^{1l}(\alpha_{s},m_{Q,bare})\bigg], \label{rn1l}
\end{eqnarray}
where the $\mathcal{A}^{il}_{bare}|_{i=0,1}$ represent the tree and one-loop bare 
amplitudes, respectively. Expanding the right-hand of Eq.(\ref{rn1l}) in series of $\alpha_s$ and keeping the results up to $\mathcal{O}(\alpha_s)$, we then express 
$\mathcal{A}(\alpha_s,m_Q)$ in terms of the renormalized quantities 
\begin{eqnarray}
\mathcal{A}(\alpha_s,m_Q)&=&\mathcal{A}^{0}(m_Q)+\mathcal{A}^{1}(\alpha_s,m_Q)+\mathcal{O}(\alpha_s^2). 
\end{eqnarray}
Now $\mathcal{A}^{i}|_{i=0,1}$ are the finite tree and one-loop renormalized amplitudes, 
respectively. Note that the IR divergences in the loop integrals cancel exactly with 
those in $\delta Z_2$. 

The analytical results of squared amplitude at QCD NLO are too complicated to present, here we only show the LO results:
\begin{subequations}\label{amp2}
\begin{align}
	|\mathcal{A}_{^1S_0^{[1]}}^0|^2 = \frac{96 e_Q^4 (4\pi\alpha)^3}{m_{12}^4 (m_{12}^2 - m_H^2)^2} \bigg[ m_{12}^6 + 2 m_{12}^4 (m_{23}^2 - m_H^2) + 2 m_l^2 m_H^4 \notag \\
	\quad + m_{12}^2 \big( 2 m_{23}^4 - 2 m_{23}^2 (2 m_l^2 + m_H^2) + 2 m_l^4 - 2 m_l^2 m_H^2 + m_H^4 \big) \bigg],  \\[10pt]
	|\mathcal{A}_{^3P_0^{[1]}}^0|^2 = \frac{128 e_Q^4 (4\pi\alpha)^3(m_{12}^2 - 3 m_H^2)^2}{m_H^2 (m_{12}^3 - m_{12} m_H^2)^4} \bigg[ m_{12}^6 + 2 m_l^2 m_H^4 + 2 m_{12}^4 (m_{23}^2 - m_H^2) \notag \\
	\quad + m_{12}^2 \big( 2 m_{23}^4 + 2 m_l^4 - 2 m_l^2 m_H^2 + m_H^4 - 2 m_{23}^2 (2 m_l^2 + m_H^2) \big) \bigg], \\[10pt]
	|\mathcal{A}_{^3P_1^{[1]}}^0|^2 = \frac{768e_Q^4 (4\pi\alpha)^3}{m_H^2 (m_{12}^2 - m_H^2)^4} \bigg[ m_{12}^6 + 2 m_{12}^4 (m_{23}^2 - m_H^2) \notag \\
	\quad + 2 m_H^2 \big( -2 m_{23}^4 + 2 m_{23}^2 (2 m_l^2 + m_H^2) - m_l^2 (2 m_l^2 + m_H^2) \big) \notag \\
	\quad + m_{12}^2 \big( 2 m_{23}^4 + 2 m_l^4 + 2 m_l^2 m_H^2 + m_H^4 - 2 m_{23}^2 (2 m_l^2 + 3 m_H^2) \big) \bigg], \\[10pt]
	|\mathcal{A}_{^3P_2^{[1]}}^0|^2 = \frac{256e_Q^4 (4\pi\alpha)^3}{m_H^2 (m_{12}^3 - m_{12} m_H^2)^4} \bigg[ m_{12}^{10} + 12 m_l^2 m_H^8 + 2 m_{12}^8 (m_{23}^2 - m_H^2) \notag \\
	\quad - 2 m_{12}^4 m_H^2 \big( 6 m_{23}^4 + 6 m_l^4 + 5 m_l^2 m_H^2 + 6 m_H^4 - 12 m_{23}^2 (m_l^2 + m_H^2) \big) \notag \\
	\quad + 6 m_{12}^2 m_H^4 \big( 2 m_{23}^4 + 2 m_l^4 - 2 m_l^2 m_H^2 + m_H^4 - 2 m_{23}^2 (2 m_l^2 + m_H^2) \big) \notag \\
	\quad + m_{12}^6 \big( 2 m_{23}^4 + 2 m_l^4 + 10 m_l^2 m_H^2 + 7 m_H^4 - 2 m_{23}^2 (2 m_l^2 + 7 m_H^2) \big) \bigg].
\end{align}
\end{subequations}

In $P$-wave cases, additional IR divergences will stem from the phase-space integration 
when the energy of the photon becomes soft, i.e. when $m_{12}$ is close to $m_H$. The 
physical origin of these IR divergences is highly analogous to that encountered in the 
inclusive light-hadron decay of $P$-wave quarkonium, where within the framework of NRQCD 
factorization, such divergences are absorbed into the CO $S$-wave 
LDMEs~\cite{Bodwin:1994jh}. In principle, to remove the IR divergence in 
$\chi_{QJ}\to l^{+}l^{-}\gamma$ process, a similar treatment can be implemented by 
incorporating the contribution of $S$-wave Fock states $|Q\bar{Q}(^3S_1^{[1]})\gamma\rangle$. Indeed, we notice that the $|Q\bar{Q}(^3S_1^{[1]})\gamma\rangle$
Fock state contribution also plays an essential role to cancel the IR divergence in the 
exclusive leptonic decay of $\chi_{QJ}\to l^{+}l^{-}$~\cite{Yang:2012gk,Kivel:2015iea,Jia:2024dzm}. 
In practice, to measure the Dalitz decay in the experiment, all the three particles must 
be identified. Therefore the energy of photon can not be too soft or it can not be 
detected.~\footnote{Private communication with Xiao-Rui Lyu from BES III Collaboration.} 
To provide a more realistic comparison between our theoretical predictions and 
experimental data, we will implement various kinematic cuts on the photon energy in the 
numerical analysis.

Substituting the expressions from eqs.~\eqref{LDMEs}, \eqref{SDCs}, and \eqref{amp2} 
into eq.~\eqref{width} and integrating out $m_{23}$, we also obtain the invariant mass
($m_{12}$) distribution of the lepton pair, which yields
\begin{subequations}
	\begin{align}
        \frac{d\Gamma_{\eta_{Q}}^{\rm LO}}{dm_{12}}&=\frac{16 e_Q^4 \alpha^3 \vert {R_S}(0)\vert^2 \sqrt{m_{12}^2 - 4 m_l^2} (m_{12}^2 + 2 m_l^2) (m_H^2-m_{12}^2)}{\pi m_{12}^4 m_{H}^4 },
   \label{loresultetac}\\
		\frac{d\Gamma_{\chi_{Q0}}^{\rm LO}}{dm_{12}} &= \frac{64 e_Q^4\alpha ^3 \lvert R_P^{\prime}(0)\rvert^2 \sqrt{m_{12}^2-4 m_l^2} \left(m_{12}^2+2 m_l^2\right) \left(m_{12}^2-3 m_H^2\right)^2}{\pi  m_{12}^4 m_H^6 \left(m_H^2-m_{12}^2\right)},\label{m12chi0} \\
		\frac{d\Gamma_{\chi_{Q1}}^{\rm LO}}{dm_{12}} &= \frac{128 e_Q^4 \alpha ^3 \lvert R_P^{\prime}(0)\rvert^2 \sqrt{m_{12}^2-4 m_l^2} \left(m_{12}^2+2 m_l^2\right) \left(m_{12}^2+m_H^2\right)}{\pi  m_{12}^2 m_H^6 \left(m_H^2-m_{12}^2\right)},\label{m12chi1} \\
		\begin{split}
			\frac{d\Gamma_{\chi_{Q2}}^{\rm LO}}{dm_{12}} &= \frac{128 e_Q^4\alpha ^3 \lvert R_P^{\prime}(0)\rvert^2 \sqrt{m_{12}^2-4 m_l^2} \left(m_{12}^2+2 m_l^2\right) \left(m_{12}^4+3m_{12}^2 m_H^2+6 m_H^4\right)}{5 \pi  m_{12}^4 m_H^6 \left(m_H^2-m_{12}^2\right)}.
		\end{split}\label{m12chi2}
	\end{align}
\end{subequations}

As we discussed above, the analytical expressions in 
Eq.(\ref{m12chi0},\ref{m12chi1},\ref{m12chi2}) explicitly demonstrate that the invariant mass distribution becomes divergent as $m_{12}$ approaches the mass of parent meson $H$.

\section{Phenomenological results} \label{II}
We first specify the input parameters used in the numerical calculations. The masses of 
the heavy quarkonia, electron, and muon are taken from the Particle Data 
Group~\cite{ParticleDataGroup:2024cfk}, which are
\begin{gather}
	m_{\eta_c} = 2.984~{\rm GeV}, \quad m_{\chi_{c0}} = 3.415~{\rm GeV}, \quad m_{\chi_{c1}} = 3.511~{\rm GeV}, \quad m_{\chi_{c2}} = 3.556~{\rm GeV}, \nonumber \\
	m_{\eta_b} = 9.399~{\rm GeV}, \quad m_{\chi_{b0}} = 9.859~{\rm GeV}, \quad m_{\chi_{b1}} = 9.893~{\rm GeV}, \quad m_{\chi_{b2}} = 9.912~{\rm GeV}, \nonumber \\
	m_{e} = 0.511 \times 10^{-3}~{\rm GeV}, \quad m_{\mu} = 0.106~{\rm GeV}.
\end{gather}
The values of $\vert R_S(0)\vert^2$ and $\vert {R'_{P}}(0)\vert^2$ are determined in
a potential model calculation with the Buchm\"{u}ller-Tye potential~\cite{Eichten:1995ch}:
\begin{align}
	\vert R_S^{\eta_c}(0)\vert^2 &= 0.810~{\rm GeV}^3, & \vert R_S^{\eta_b}(0)\vert^2 &= 6.477~{\rm GeV}^3, \nonumber \\
	\vert {R'_{P}}^{\chi_{cJ}}(0)\vert^2 &= 0.075~{\rm GeV}^5, & \vert {R'_{P}}^{\chi_{bJ}}(0)\vert^2 &= 1.417~{\rm GeV}^5.
\end{align}
The QED coupling is fixed at $\alpha=1/137$, while the world averaged value 
$\alpha_s(m_Z)=0.1180$~\cite{ParticleDataGroup:2024cfk} is adopted for the strong 
coupling constant. We employ the \textsf{RunDec3} package~\cite{Herren:2017osy} 
to evaluate $\alpha_s(\mu_R)$ from starting point $\alpha_s(m_Z)$ with one-loop 
running, where the renormalization scale is chosen as $\mu_R = m_H$.

\begin{figure}[htbp]
	\centering
	\includegraphics[width=0.45\textwidth]{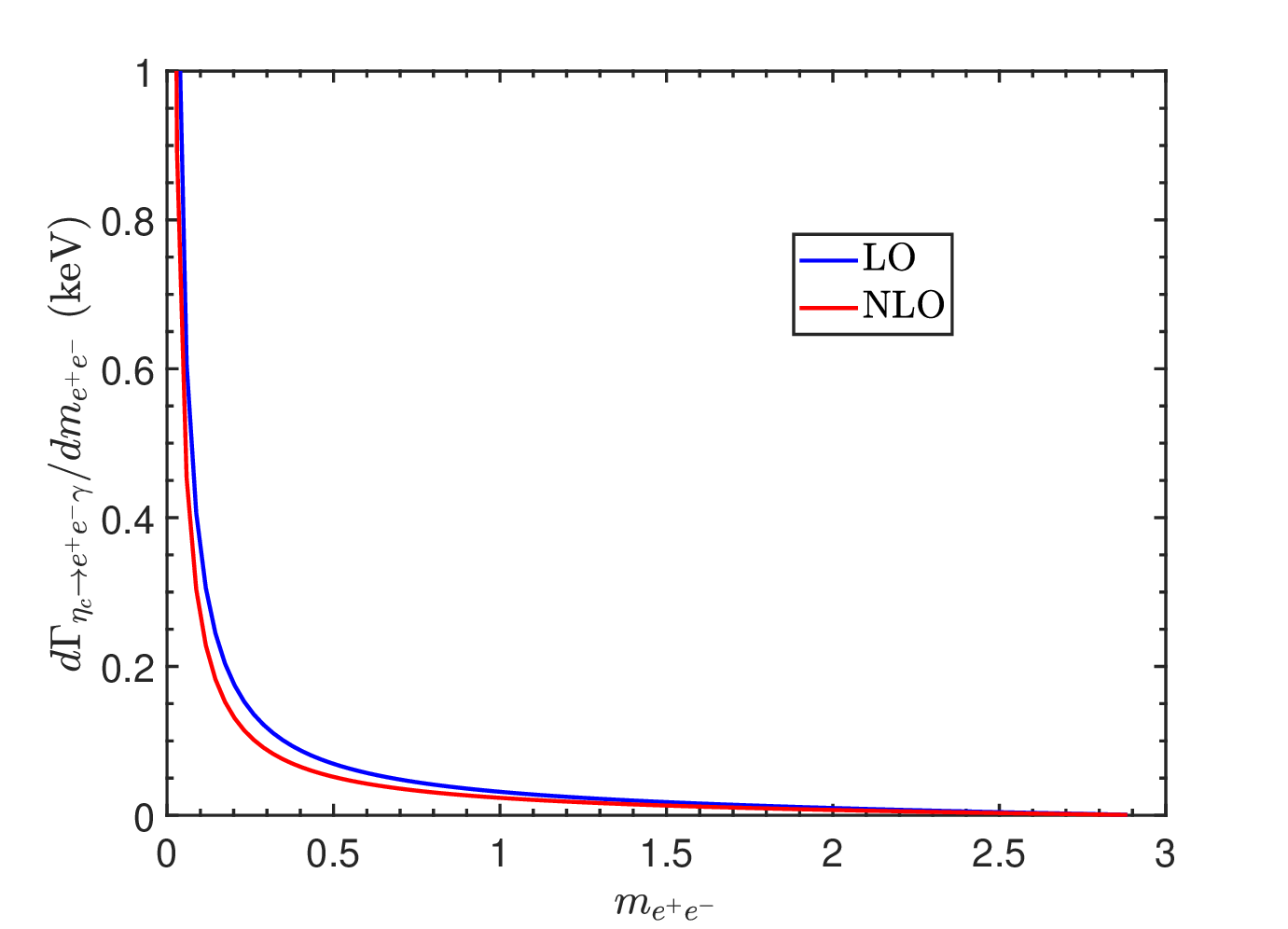}
	\includegraphics[width=0.45\textwidth]{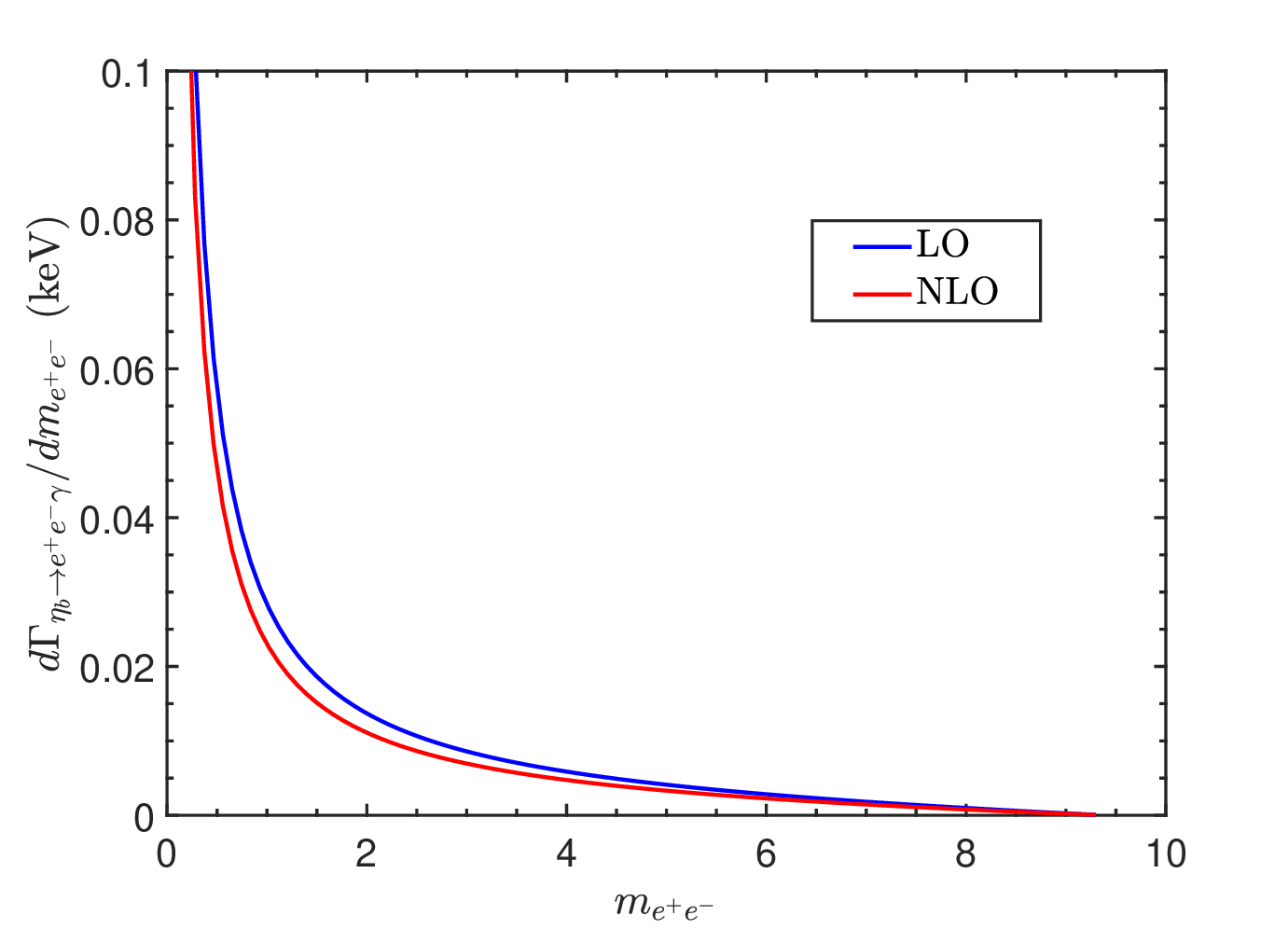}
	\includegraphics[width=0.45\textwidth]{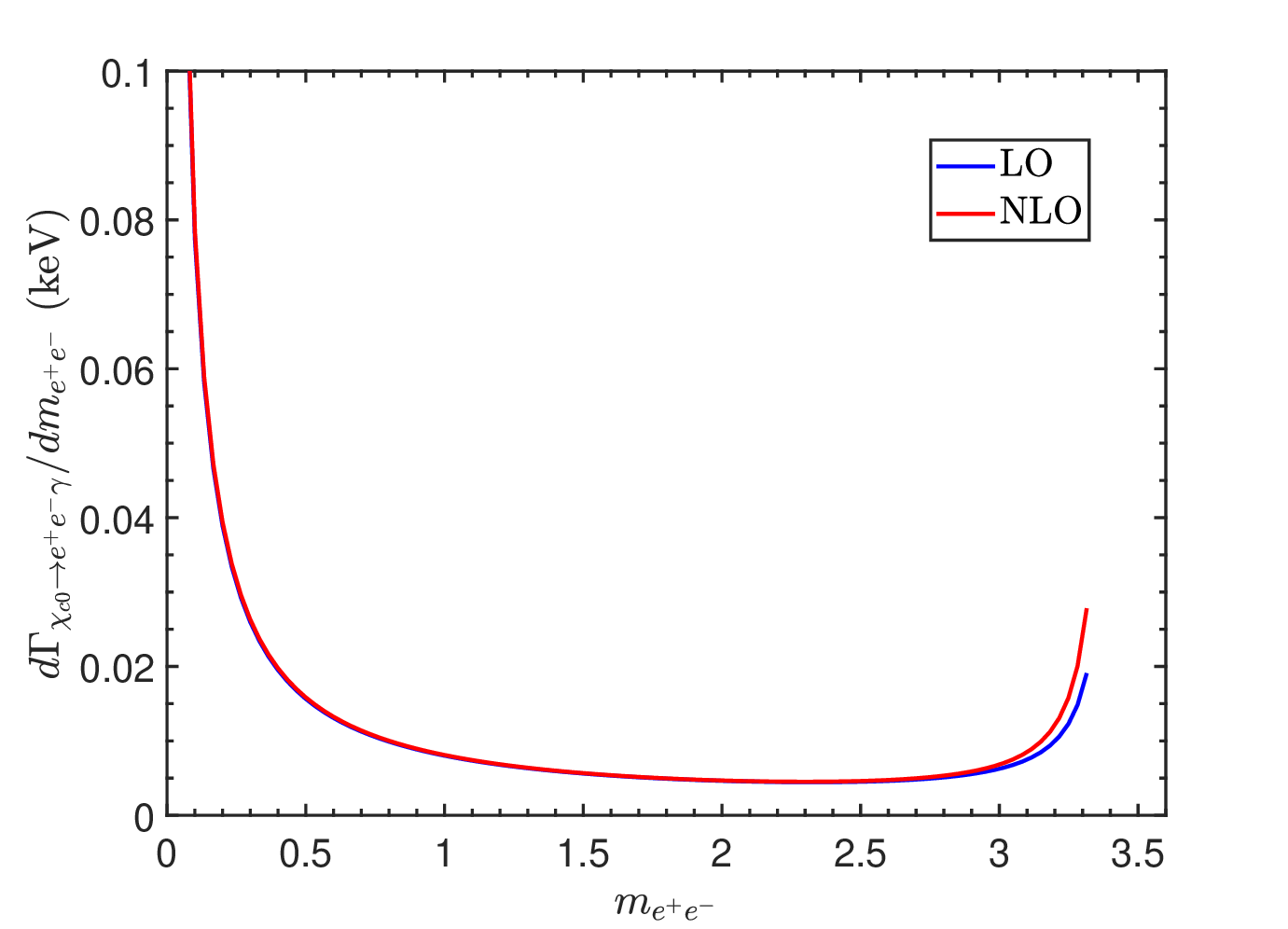}
	\includegraphics[width=0.45\textwidth]{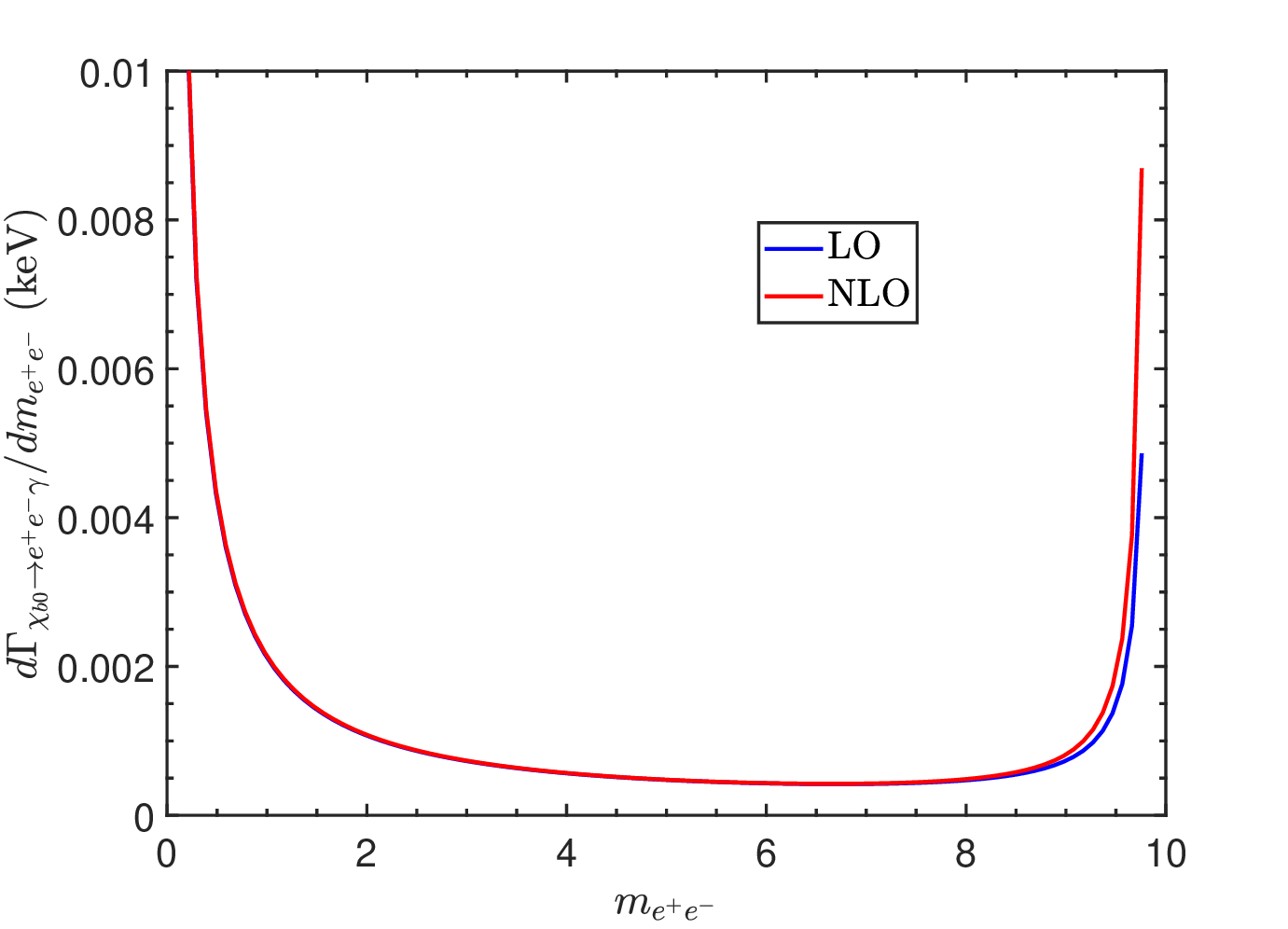}
	\includegraphics[width=0.45\textwidth]{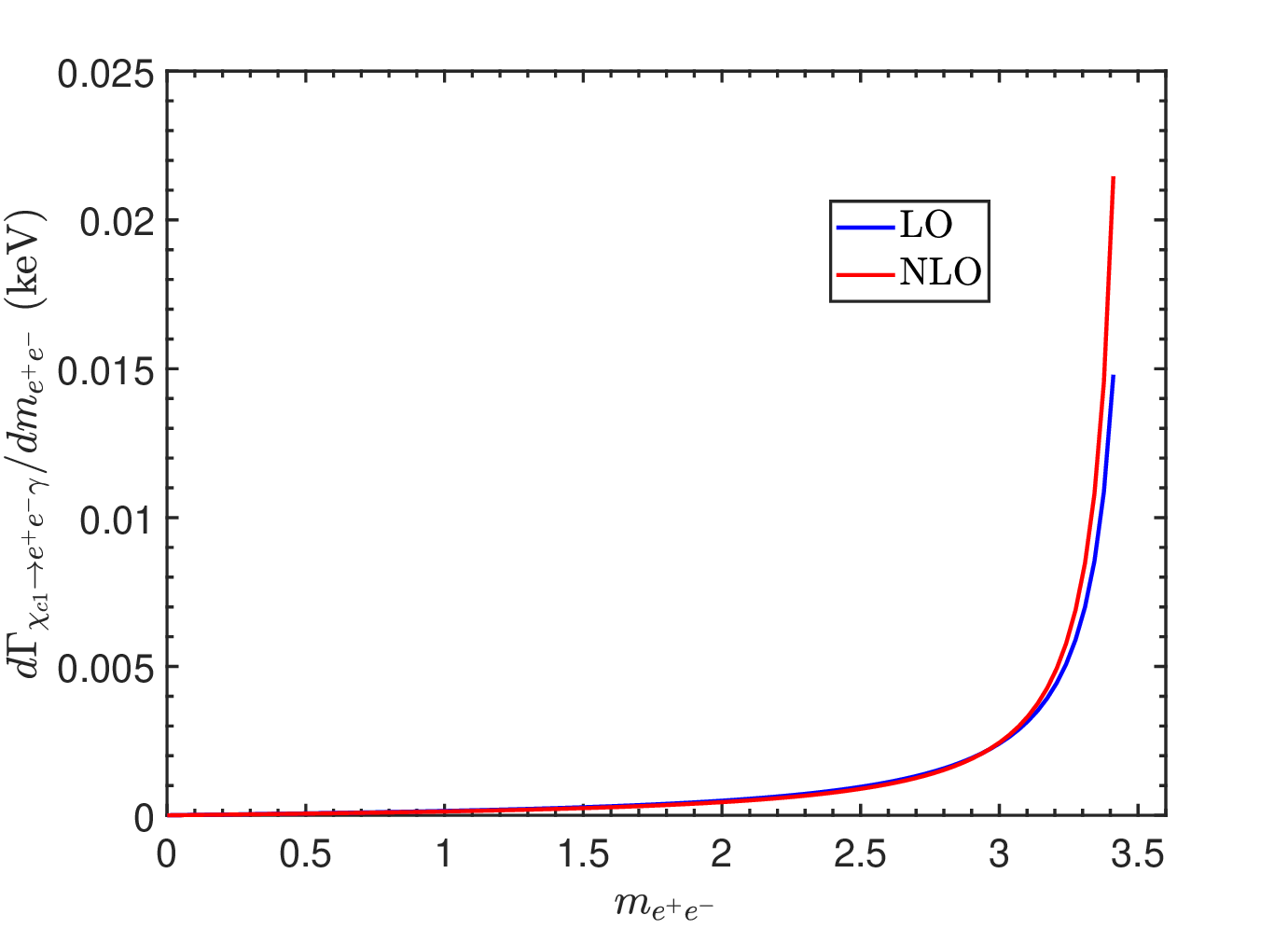}
	\includegraphics[width=0.45\textwidth]{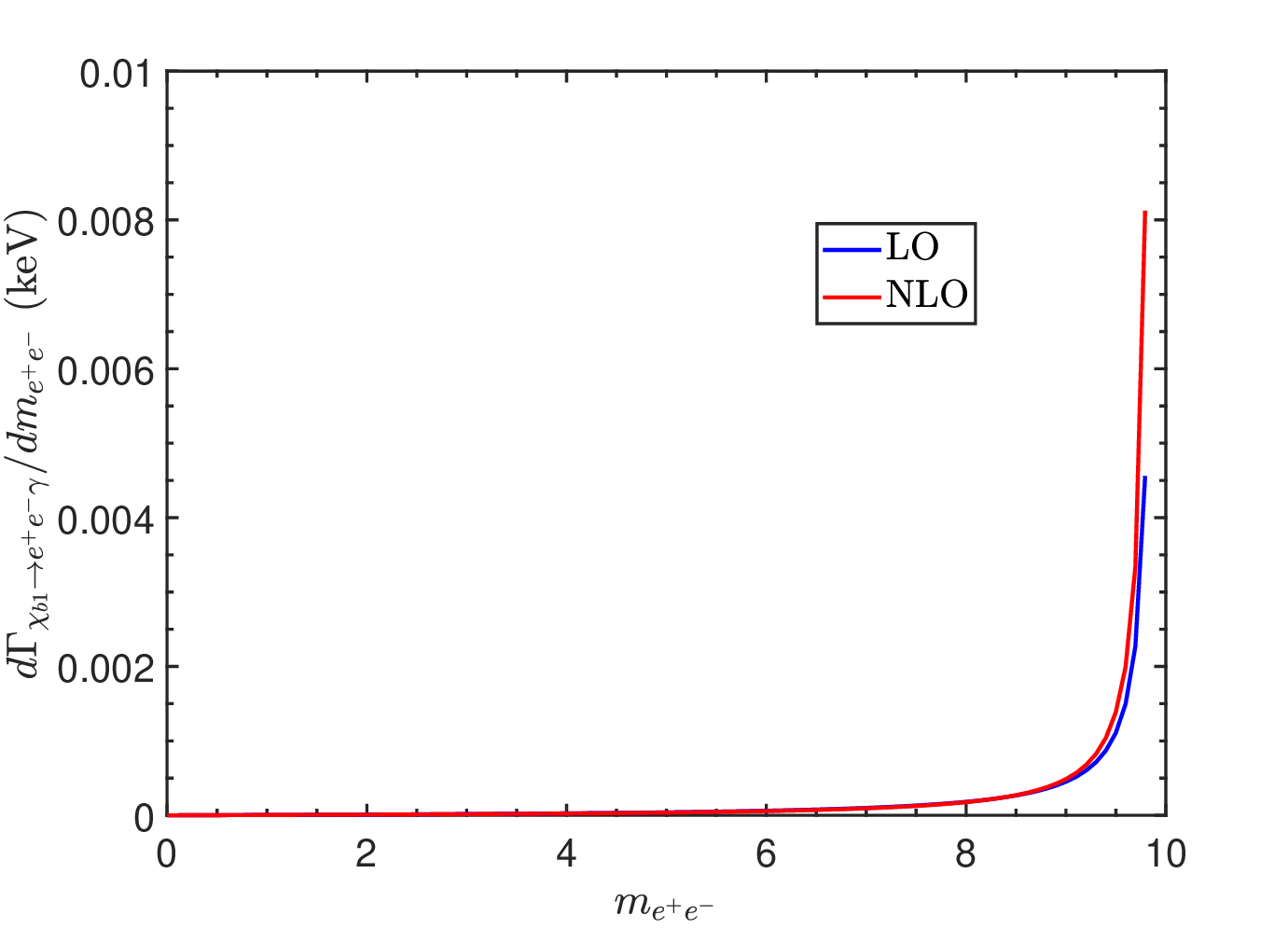}
	\includegraphics[width=0.45\textwidth]{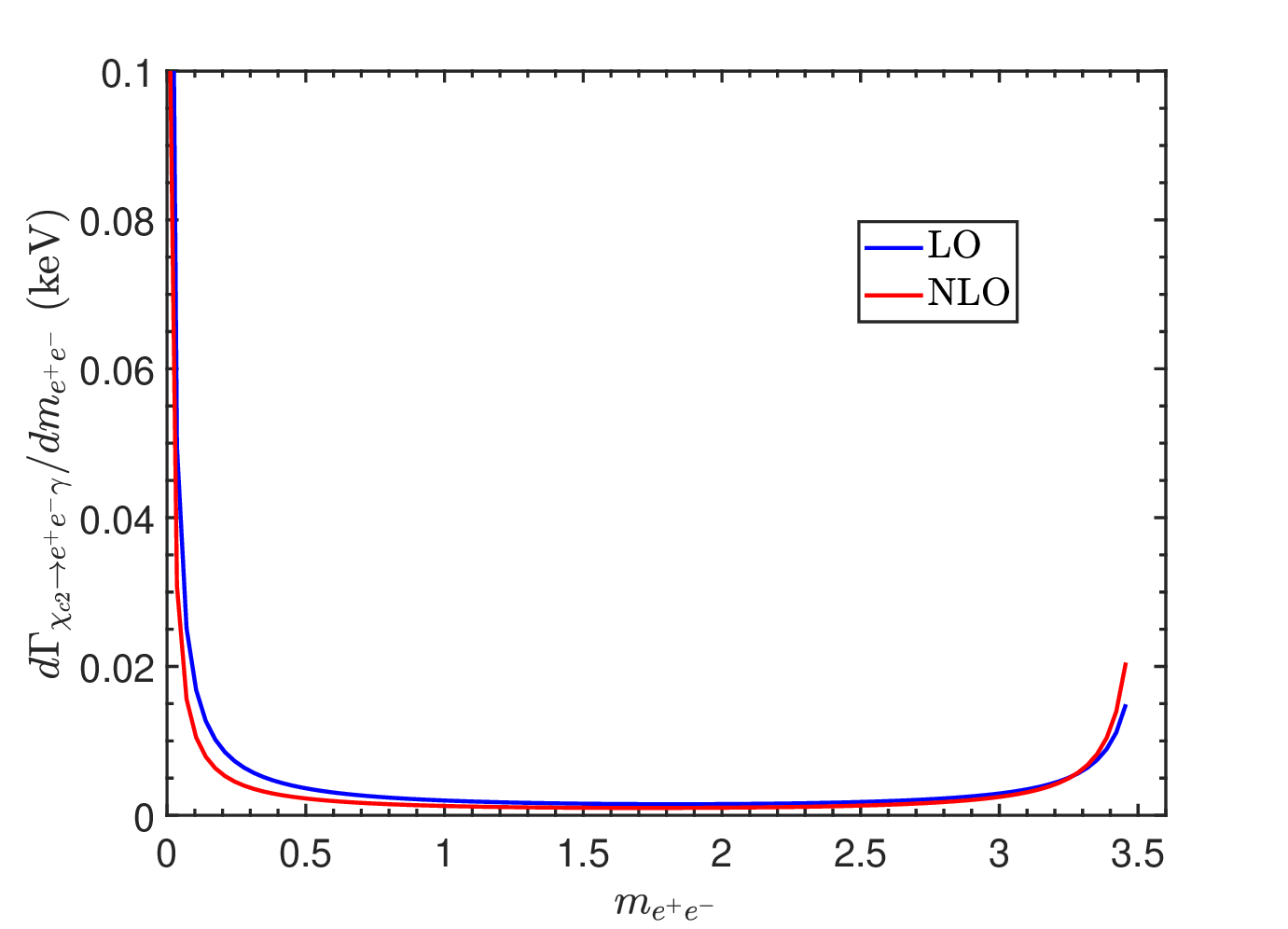}
	\includegraphics[width=0.45\textwidth]{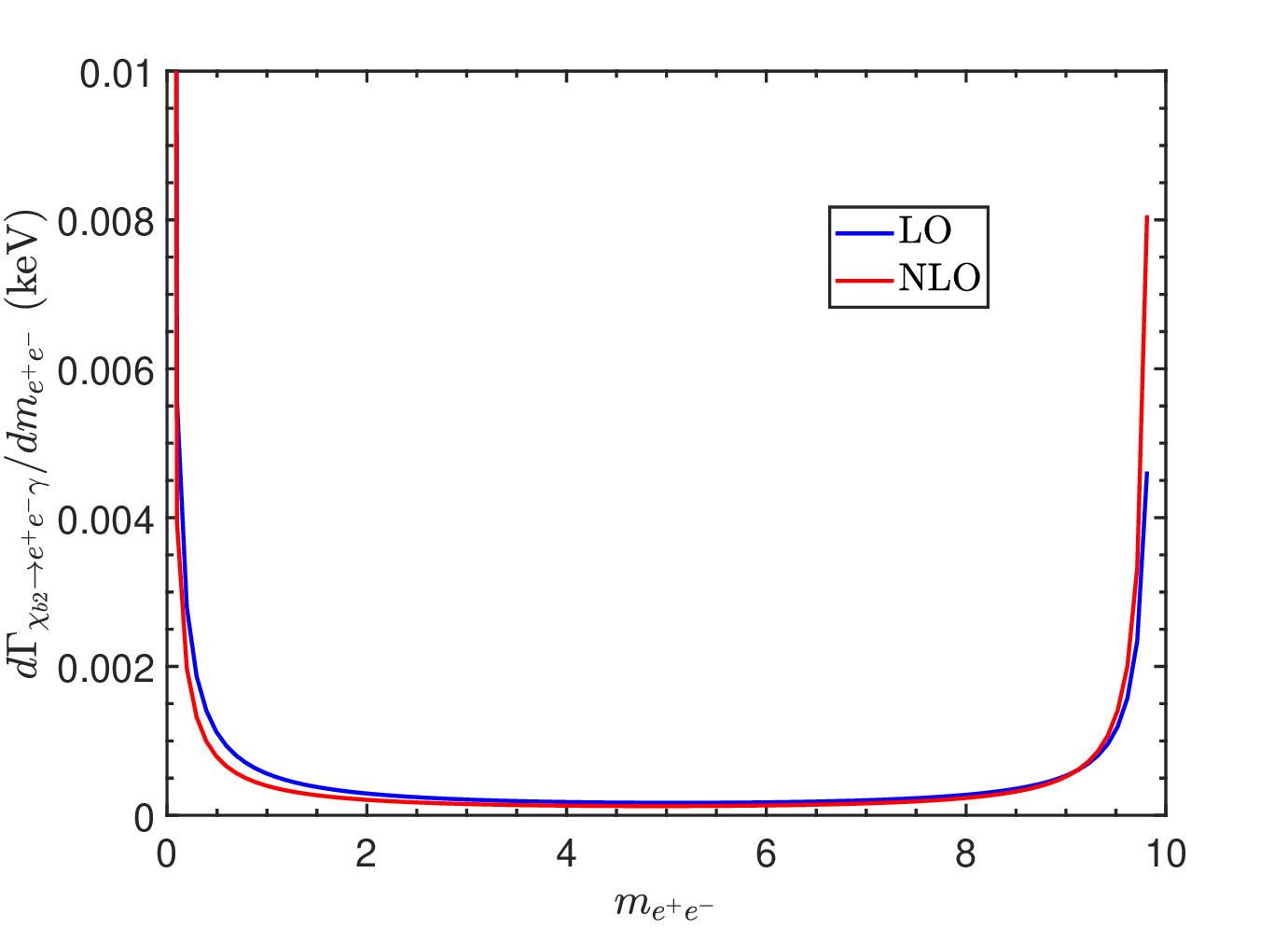}
	\caption{The differential decay widths $d\Gamma_{H \to e^{+}e^{-}\gamma}/dm_{e^+e^-}$ as functions of the $e^{+}e^{-}$ invariant mass $m_{e^+e^-}$ at QCD LO and NLO accuracy for $H = \eta_c$, $\eta_b$, $\chi_{cJ}$ ($J=0,1,2$), and $\chi_{bJ}$ ($J=0,1,2$), respectively. The renormalization scale is chosen as $\mu_R = m_H$.} \label{Qee}
\end{figure}

\begin{figure}[htbp]
	\centering
	\includegraphics[width=0.45\textwidth]{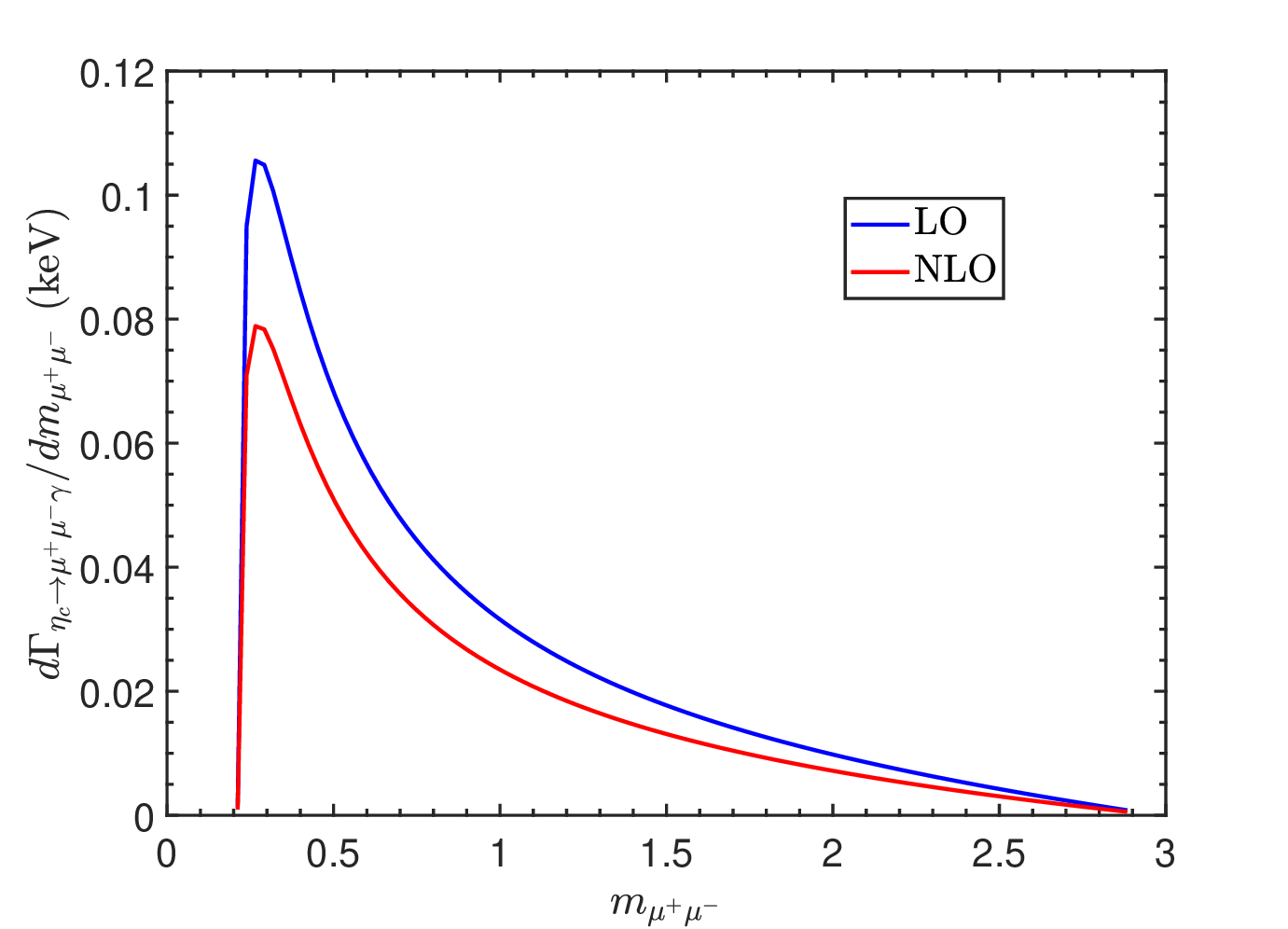}
	\includegraphics[width=0.45\textwidth]{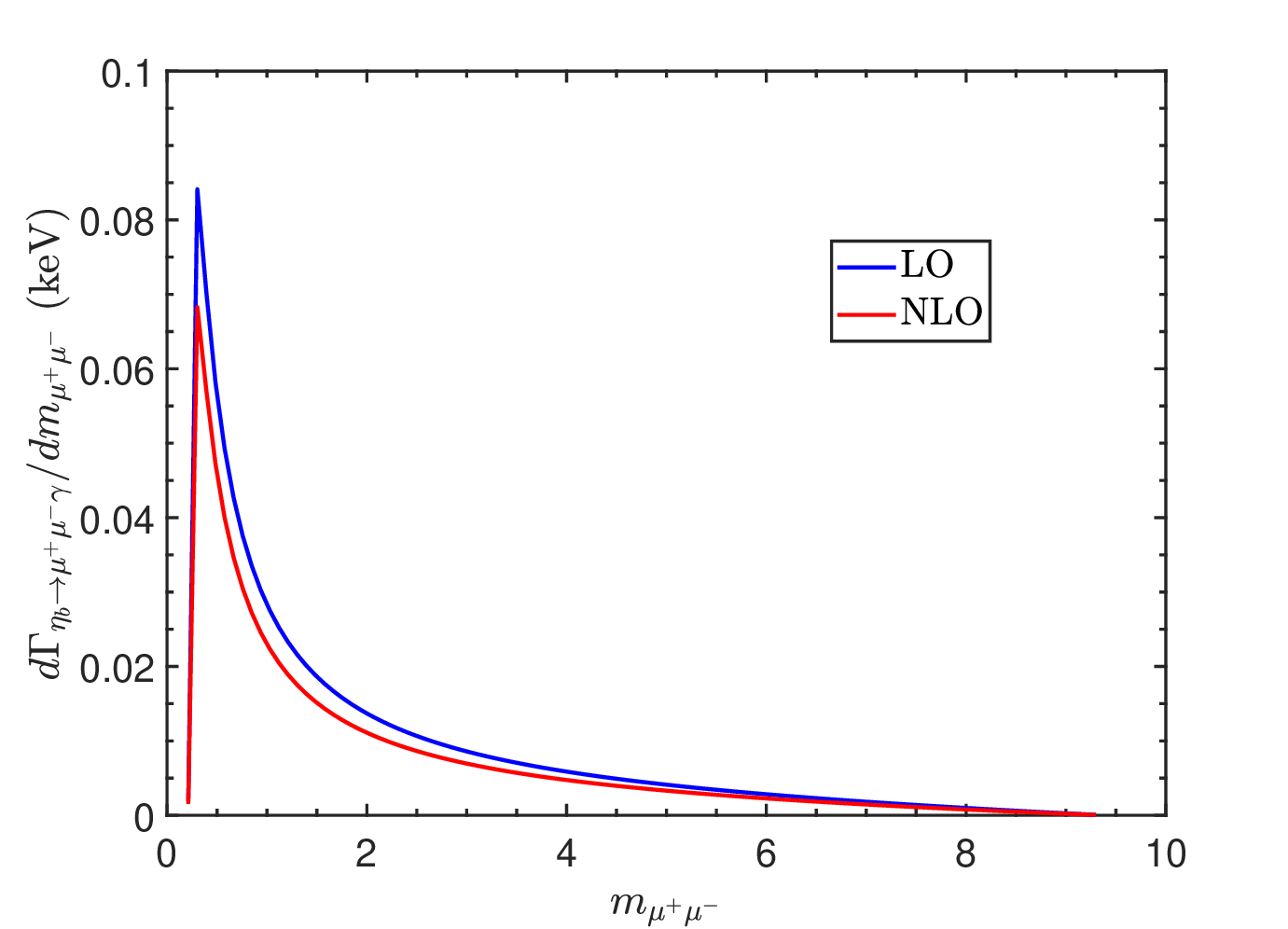}
	\includegraphics[width=0.45\textwidth]{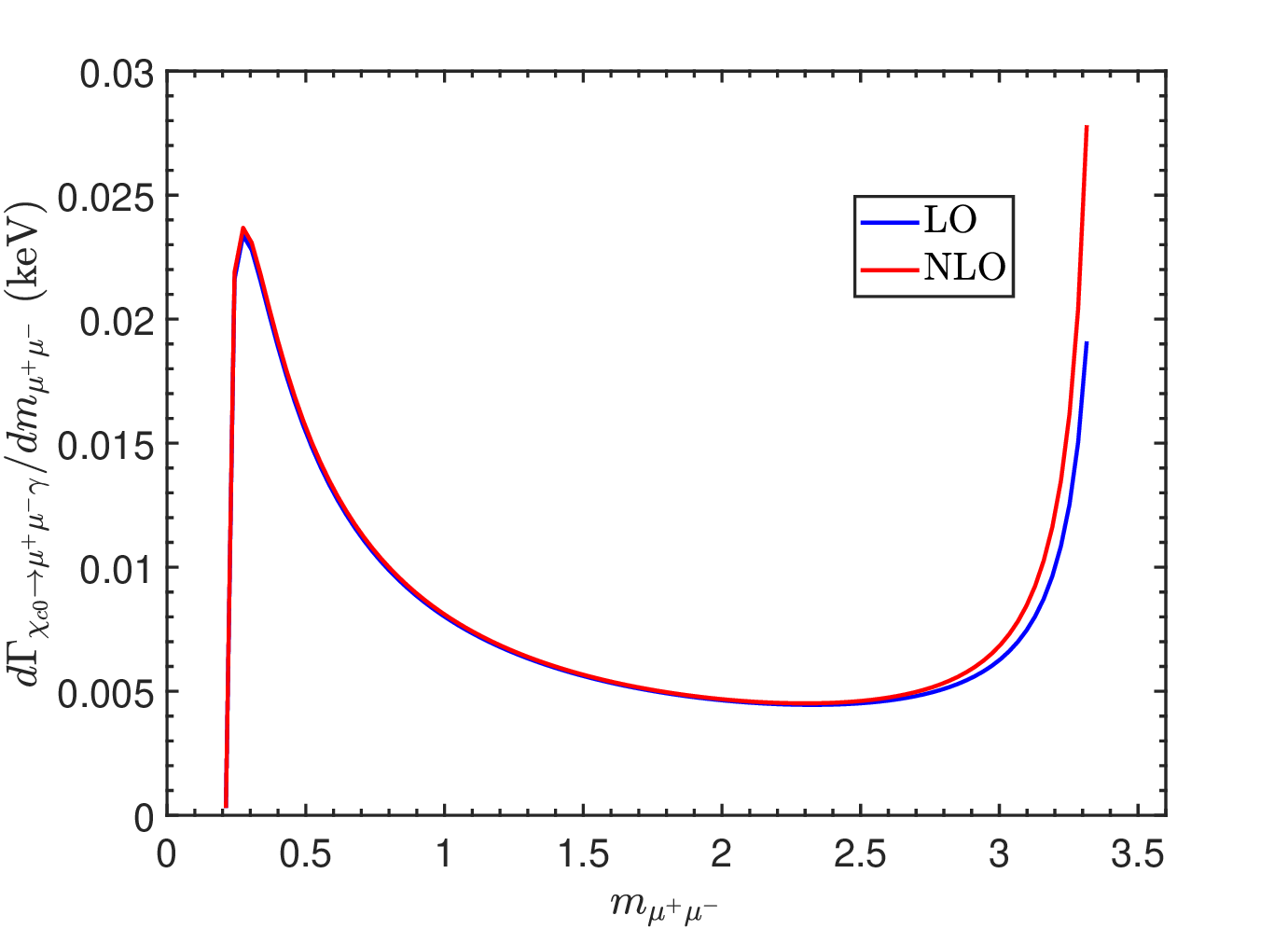}
	\includegraphics[width=0.45\textwidth]{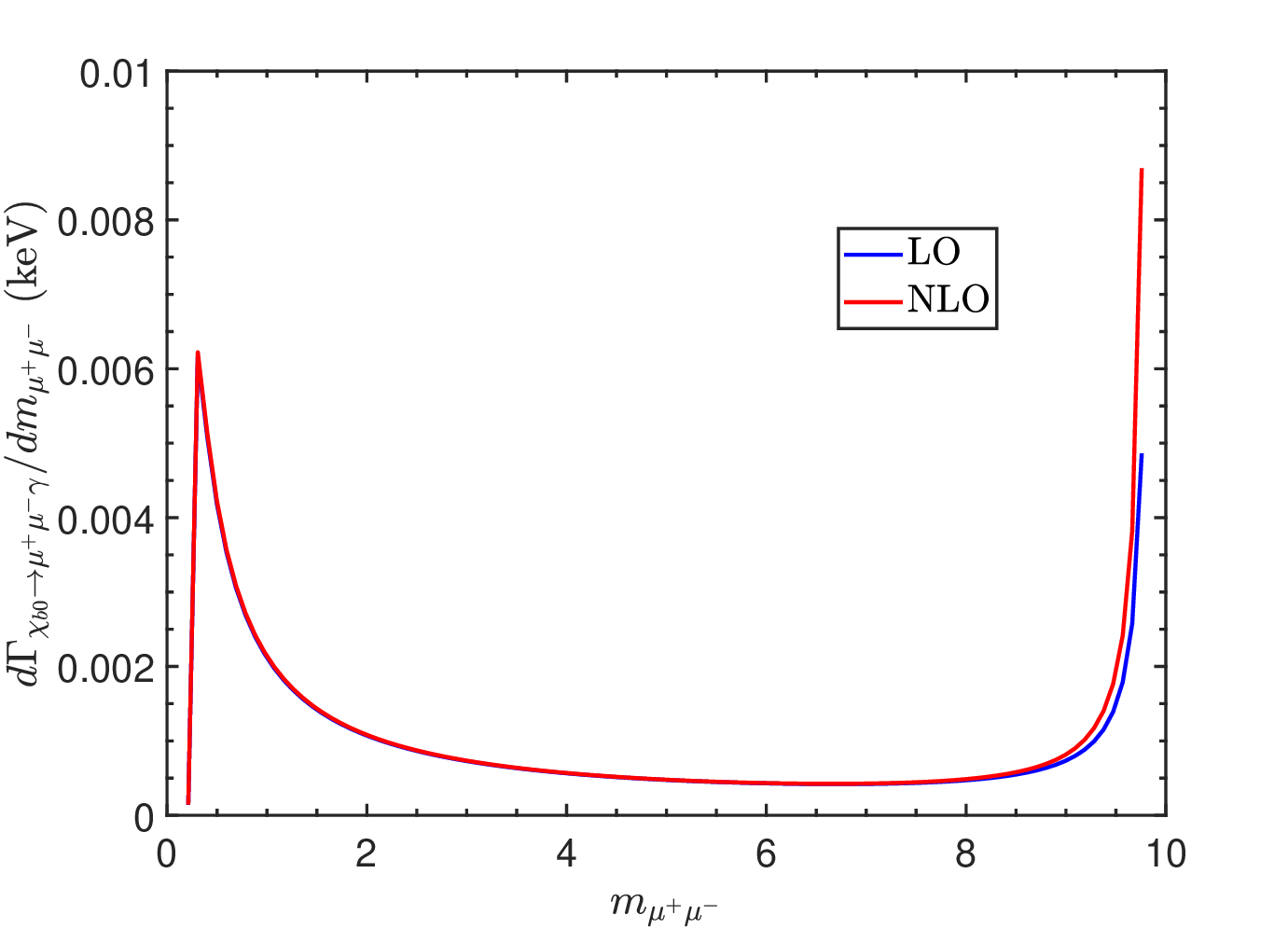}
	\includegraphics[width=0.45\textwidth]{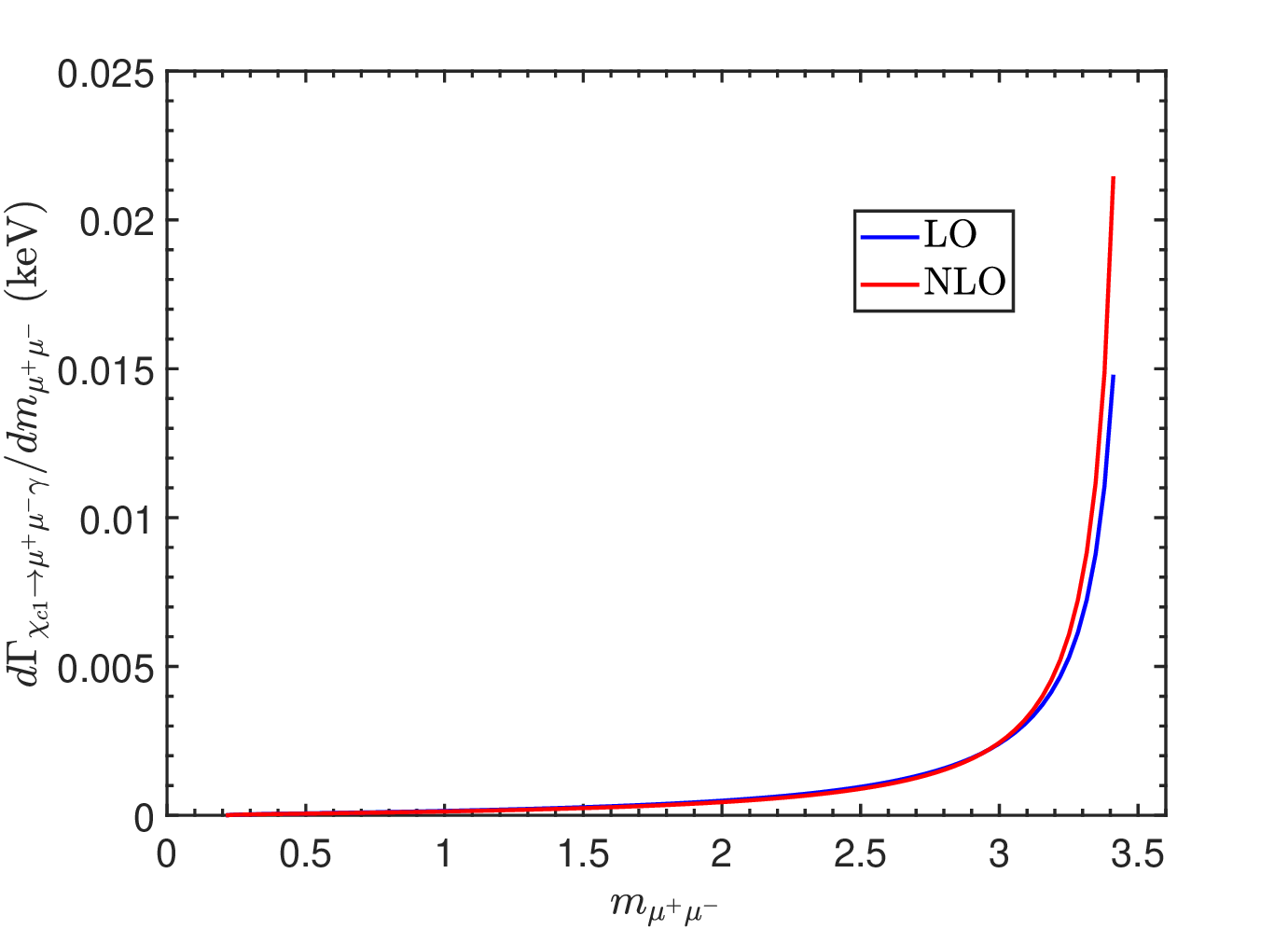}
	\includegraphics[width=0.45\textwidth]{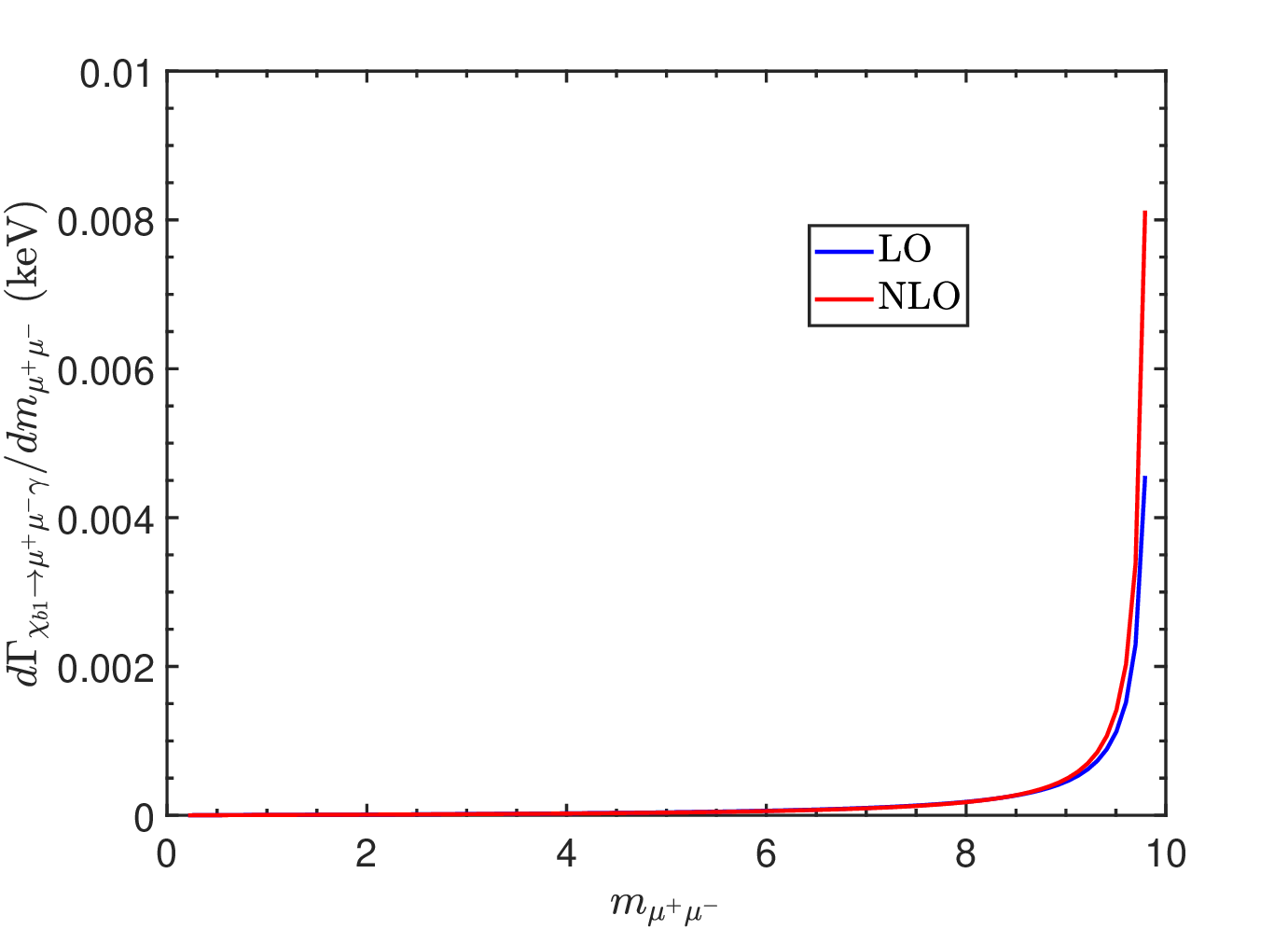}
	\includegraphics[width=0.45\textwidth]{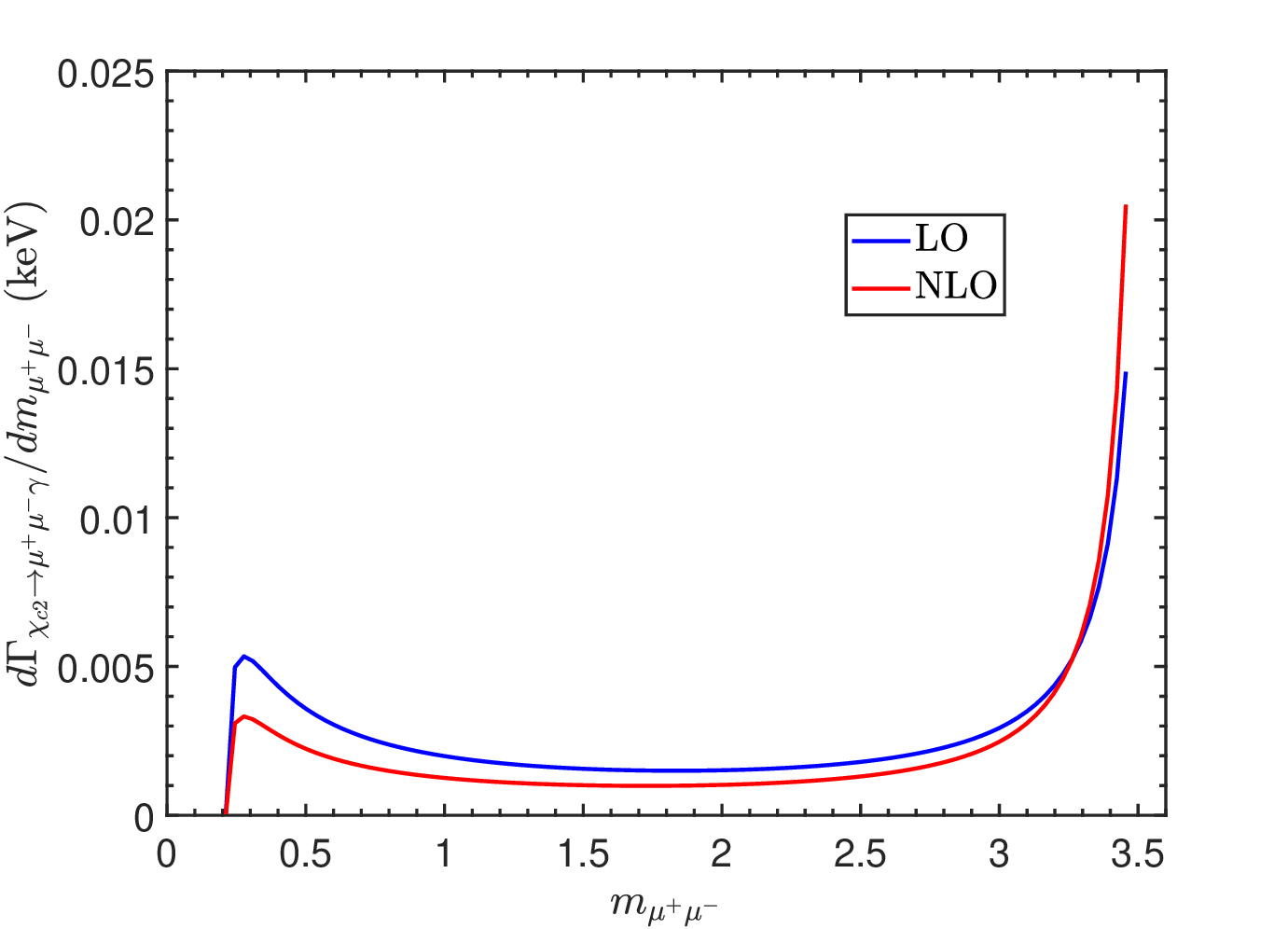}
	\includegraphics[width=0.45\textwidth]{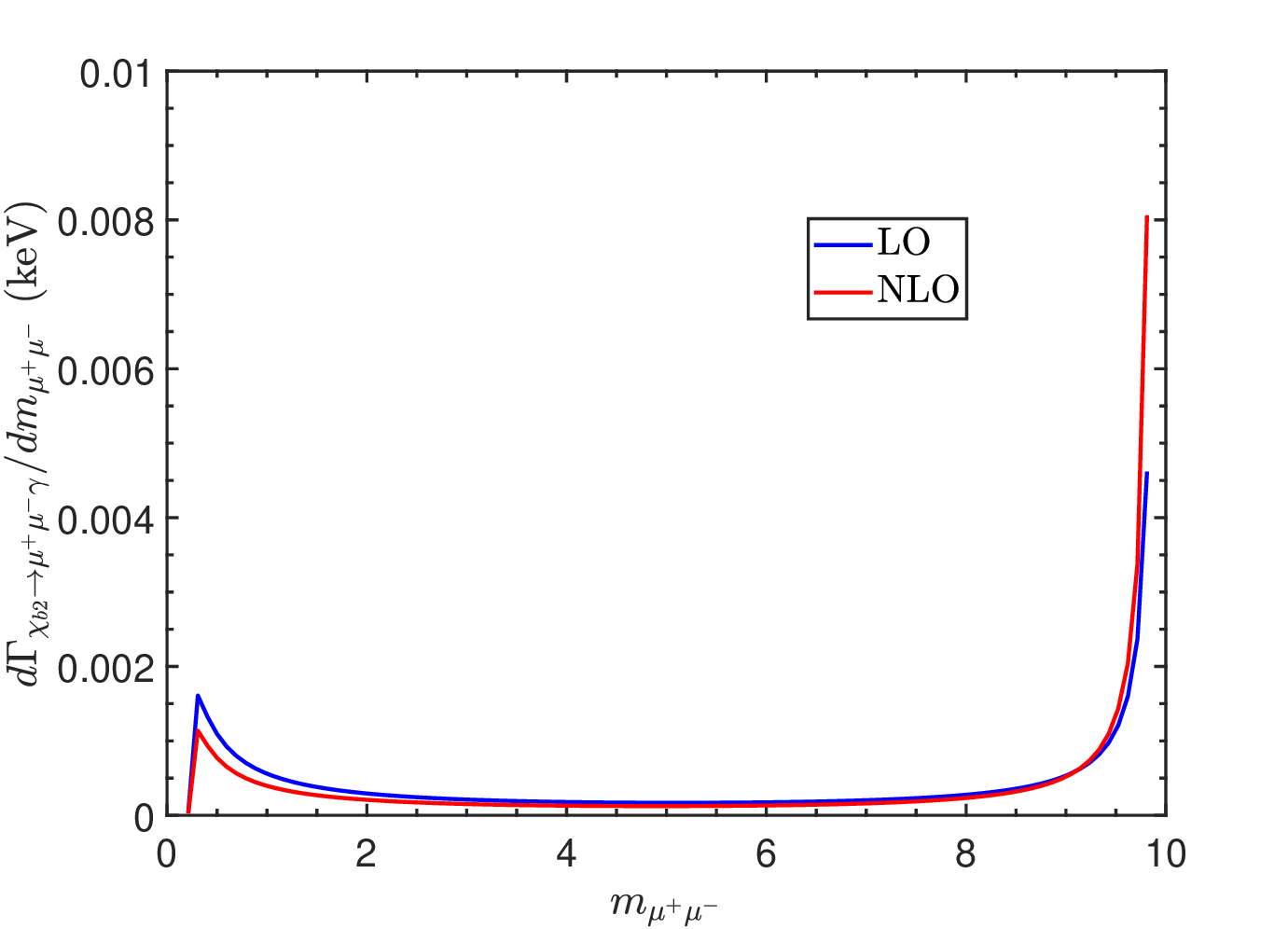}
	\caption{The differential decay widths $d\Gamma_{H \to \mu^{+}\mu^{-}\gamma}/dm_{\mu^+ \mu^-}$ as functions of the $\mu^{+}\mu^{-}$ invariant mass $m_{\mu^+ \mu^-}$ at QCD LO and NLO accuracy for $H = \eta_c$, $\eta_b$, $\chi_{cJ}$ ($J=0,1,2$), and $\chi_{bJ}$ ($J=0,1,2$), respectively. The renormalization scale is chosen as $\mu_R = m_H$} \label{Qmumu}
\end{figure}

Utilizing these input parameters, we first evaluate the LO and NLO differential decay 
widths with respect to the dilepton invariant mass, $d\Gamma_{H\to l^{+}l^{-}\gamma}/dm_{12}$, spanning the physical phase space 
$m_{12} \in [2m_l,\sqrt{m_H^2-2 m_H E_{\gamma}}]$ with $E_{\gamma}=100~\rm{MeV}$. We 
present the results for electron and muon channels in Fig.~\ref{Qee} and ~\ref{Qmumu}, 
respectively. It can be clearly seen that except for the axial-vector state 
$\chi_{Q1}$, distinct peaks emerge near the lepton-pair threshold, which correspond to 
the hard collinear singularities in the massless limit. This behavior is fully 
consistent with the Landau-Yang theorem, which strictly prohibits a spin-1 particle 
from decaying into two real photons, thereby eliminating the threshold collinear 
divergence exclusively for the $\chi_{Q1}$ states. For the remaining states, the 
physical regularization role of the lepton mass is perfectly demonstrated by 
contrasting the two spectra: the dimuon channels in fig.~\ref{Qmumu} exhibit 
significantly less pronounced and more heavily suppressed singular peaks near 
$m_{12} \sim 2m_\mu$ than the dielectron modes in fig.~\ref{Qee}. This damping is a 
direct consequence of the larger muon mass ($m_\mu \gg m_e$) shielding the virtual 
photon propagator pole ($p_{\gamma^*}^2 = m_{12}^2 \approx 0$) and smoothing the 
distribution across the low-invariant-mass threshold.

Notably, an additional soft-photon infrared singularity emerges for the $P$-wave states 
$\chi_{cJ}$ and $\chi_{bJ}$ near the upper phase-space boundary, $m_{12} \to m_H$, 
as in the previous section we have addressed that the NRQCD factorization breaks down 
when the photon energy becomes soft. Theoretically, to obtain reliable predictions of 
the total decay widths the contribution of the Fock states 
$|Q\bar{Q}(^3S_1^{[1]})\gamma\rangle$ is needed. However, for the differential distribution, it is not known quantitatively below which exact value of the photon energy the theoretical predictions will become unreliable. Therefore, the Dalitz decays of the $P$-wave $\chi_{QJ}$ states provide a unique opportunity to test our understanding of the theory near this threshold region.

To compare with the experimental measurements, we discretely consider three different cuts on the photon energy, i.e., $E^{\rm cut}_{\gamma}>100$, $300$, and $500~\text{MeV}$. The total decay widths at QCD LO and NLO for the different cuts are listed in table~1 and table~2 for $l=e$ and $l=\mu$, respectively. We find that 
for $S$-wave states $\eta_{c,b}$, the theoretical predictions remain almost unchanged 
under variation of these cuts and the NLO QCD corrections are negative. Specifically, 
they can lead to roughly $25\%$ reduction for $\eta_{c}\to l^{+}l^{-}\gamma$ and $19\%$ 
for $\eta_{b}\to l^{+}l^{-}\gamma$. Our LO results agree with those in 
Ref.~\cite{Jia:2009ip}, and our central QCD NLO predictions are 
\begin{equation}
	\begin{aligned}
		\Gamma(\eta_{c}\to e^{+}e^{-}\gamma) = 195~\text{eV},\quad
		\Gamma(\eta_{c}\to \mu^{+}\mu^{-}\gamma) = 53.1~\text{eV}&\\
		\Gamma(\eta_{b}\to e^{+}e^{-}\gamma) = 12.3~\text{eV},\quad
\Gamma(\eta_{b}\to \mu^{+}\mu^{-}\gamma) = 4.58~\text{eV}		
	\end{aligned}
\end{equation}

Compared to $\eta_{Q}$ mesons, the $P$-wave $\chi_{QJ}$ Dalitz decays exhibit a 
significantly more complex sensitivity to the photon energy cuts. Theoretical 
predictions in these scenarios depend not only on the total angular momentum $J$ but 
also tie intrinsically to the flavors of heavy quarks and leptons. As 
$E_{\gamma}^{\rm cut}$ increases from $100~\text{MeV}$ to $500~\text{MeV}$ in the 
electron channel shown in Table~\ref{gammaee}, the LO decay widths decrease by $5\%$, 
$60\%$, and $14\%$ for the charmonium states ($\chi_{c0},\chi_{c1},\chi_{c2}$), and by 
$4\%$, $50\%$, and $12\%$ for the bottomonium states ($\chi_{b0},\chi_{b1},\chi_{b2}$),
respectively. At NLO accuracy, these reductions become even more pronounced, escalating 
to $6\%$, $66\%$, and $21\%$ for the $\chi_{cJ}$ sector, and $5\%$, $50\%$, and $18\%$
for the $\chi_{bJ}$ sector. This pattern explicitly reveals an enhanced sensitivity to 
the photon energy cut at higher perturbative orders, where the NLO width reductions 
outpace their LO counterparts by up to $10\%$ across all $\chi_{QJ}$ ($Q=c,b$)
configurations. This systematic amplification indicates that virtual loop corrections at the QCD NLO make the singular behavior near the soft-photon region more pronounced.

However, in the muon modes, theoretical predictions for the $J=0$ and $J=2$ states 
exhibit a significantly higher sensitivity to the photon energy cut; whereas the 
$\chi_{Q1}$ states yield almost identical decay widths across all choices of 
$E_{\gamma}^{\rm cut}$ between the electron and muon channels. This is due to the fact 
that for the $\chi_{Q0}$ and $\chi_{Q2}$ states the predominant contribution stems not 
only from the soft-photon region but also from the hard-photon region as shown in 
Fig.~\ref{Qee} and \ref{Qmumu}, where the two leptons are nearly collinear and the 
singularity is regularized by the lepton mass. In contrast, this collinear contribution 
almost vanishes for the $\chi_{Q1}$ decays as a consequence of the Landau-Yang theorem. 
Quantitatively, within the muon sector, when $E_{\gamma}^{\rm cut}$ increases from 
$100~\text{MeV}$ to $500~\text{MeV}$, as shown in Table~\ref{gammamumu}, the baseline LO decay widths are suppressed by $15\%$ ($\chi_{c0}$), $60\%$ ($\chi_{c1}$), and $28\%$ ($\chi_{c2}$) for the charmonium states, while the corresponding suppressions of 
$8\%(\chi_{b0})$, $50\%(\chi_{b1})$, and $21\%(\chi_{b2})$ are induced across the
bottomonium multiplet. Moving to NLO accuracy further expands this sensitivity window, 
with the percentage drops moving up to $17\%$, $66\%$, and $39\%$ for the $\chi_{cJ}$ 
states, and $11\%$, $50\%$, and $29\%$ for the $\chi_{bJ}$ states, respectively.

Beyond the cut dependence, the effects of NLO QCD radiative corrections also exhibit a 
significant diversity, which indicates that no simple, global $K$-factor can be applied 
to $P$-wave quarkonium Dalitz decays. While the axial-vector $\chi_{Q1}$ states receive NLO corrections that can be positive or negative depending on the cutoff, the tensor states suffer from severe 
negative NLO corrections. Meanwhile, for the scalar states ($\chi_{c0}$ and $\chi_{b0}$), the NLO QCD corrections are less than $6\%$, making the theoretical predictions remarkably stable. For instance, at the baseline cut of $E_{\gamma}^{\rm cut} = 100~\text{MeV}$, the NLO corrections shift the total decay widths $\Gamma_{H \to e^+e^-\gamma}$ by $2.4\%$ ($\chi_{c0}$), $9.9\%$ ($\chi_{c1}$), $-30\% (\chi_{c2})$, $3.1\%$ ($\chi_{b0}$), $20\%$ ($\chi_{b1}$) and $-19\%(\chi_{b2})$. In the $H \to \mu^+\mu^-\gamma$ decays, the corresponding NLO corrections shift the total widths by $4.4\%$ ($\chi_{c0}$), $9.9\%$ ($\chi_{c1}$), $-21.0\%$ ($\chi_{c2}$), $5.1\%$ ($\chi_{b0}$), $20\%$ ($\chi_{b1}$), and $-12.5\%$ ($\chi_{b2}$). 

We further systematically investigate the theoretical uncertainties by varying the renormalization scale $\mu_R$ from $m_H/2$ to $2m_H$. For the electron decay channel at $E_{\gamma}^{\rm cut}=100~\rm MeV$, we find the scale uncertainties are tightly bounded at $16\%$ ($\eta_c$), $7\%$ ($\eta_b$), $1\%$ ($\chi_{c0}$), $4\%$ ($\chi_{c1}$), $18\%$ ($\chi_{c2}$), $1\%$ ($\chi_{b0}$), $5\%$ ($\chi_{b1}$), and $7\%$ ($\chi_{b2}$). In the muon channel, these uncertainties shift to $16\%$ ($\eta_c$), and $7\%$ ($\eta_b$), $2\%$ ($\chi_{c0}$), $4\%$ ($\chi_{c1}$), $11\%$ ($\chi_{c2}$), $2\%$ ($\chi_{b0}$), $5\%$ ($\chi_{b1}$), and $3\%$ ($\chi_{b2}$). 

\begin{table}[t] 
	\centering
	\small
	\begin{tabular}{>{\centering\arraybackslash}m{1.5cm} *{8}{>{\centering\arraybackslash}p{1.2cm}}}
		\toprule
		\diagbox[width=2.0cm, height=1.2cm]{ $E_{\gamma}^{\rm cut}$ }{$\Gamma$ (\text{eV})}{$H$}& $\eta_c$ & $\chi_{c0}$ & $\chi_{c1}$ & $\chi_{c2}$ & $\eta_b$ & $\chi_{b0}$ & $\chi_{b1}$ & $\chi_{b2}$   \\
		\midrule
		100~MeV & 195 (261) & 67.8 (66.2) & 4.09 (3.72) & 12.8 (18.2) & 12.3 (15.2) & 1.34 (1.30) & 0.12 (0.10) & 0.34 (0.42)  \\
		\addlinespace
		300~MeV & 195 (261) & 64.9 (63.9) & 2.10 (2.15) & 10.8 (16.5) & 12.3 (15.2) & 1.28 (1.26) & 0.07 (0.07) & 0.30 (0.39)  \\
		\addlinespace
		500~MeV & 194 (260) & 63.4 (62.6) & 1.40 (1.49) & 10.1 (15.7) & 12.3 (15.2) & 1.27 (1.25) & 0.06 (0.05) & 0.28 (0.37)  \\
		\bottomrule
	\end{tabular}
	\caption{NRQCD predictions for the decay widths $\Gamma_{H \to e^{+} e^{-} \gamma}$ at NLO (LO) accuracy, where $H$ denotes $\eta_c$, $\chi_{cJ}$ ($J=0,1,2$), $\eta_b$, and $\chi_{bJ}$ ($J=0,1,2$). The soft-photon energy cuts are chosen as $E_{\gamma}^{\rm cut} = 100$, $300$, and $500$~MeV.}  
	\label{gammaee}
\end{table}

\begin{table}[t] 
	\centering
	\small
	\begin{tabular}{>{\centering\arraybackslash}m{1.5cm} *{8}{>{\centering\arraybackslash}p{1.2cm}}}
		\toprule
		\diagbox[width=2.0cm, height=1.2cm]{ $E_{\gamma}^{\rm cut}$ }{$\Gamma$ (\text{eV})}{$H$}& $\eta_c$ & $\chi_{c0}$ & $\chi_{c1}$ & $\chi_{c2}$ & $\eta_b$ & $\chi_{b0}$ & $\chi_{b1}$ & $\chi_{b2}$   \\
		\midrule
		100~MeV & 53.1 (71.4) & 25.9 (24.8) & 4.09 (3.72) & 6.93 (8.77) & 4.58 (5.65) & 0.62 (0.59) & 0.12 (0.10) & 0.21 (0.24)  \\
		\addlinespace
		300~MeV & 52.9 (71.1) & 22.9 (22.5) & 2.10 (2.15) & 4.99 (7.11) & 4.58 (5.65) & 0.57 (0.56) & 0.07 (0.07) & 0.17 (0.20)  \\
		\addlinespace
		500~MeV & 52.4 (70.4) & 21.5 (21.2) & 1.39 (1.48) & 4.26 (6.31) & 4.57 (5.64) & 0.55 (0.54) & 0.06 (0.05) & 0.15 (0.19)  \\
		\bottomrule
	\end{tabular}
	\caption{NRQCD predictions for the decay widths $\Gamma_{H \to \mu^{+} \mu^{-} \gamma}$ at NLO (LO) accuracy, where $H$ denotes $\eta_c$, $\chi_{cJ}$ ($J=0,1,2$), $\eta_b$, and $\chi_{bJ}$ ($J=0,1,2$). The soft-photon energy cuts are chosen as $E_{\gamma}^{\rm cut} = 100$, $300$, and $500$~MeV.}  
	\label{gammamumu}
\end{table}

\section{Summary} \label{III}

In this paper, we have performed a systematic calculation of the total and differential 
decay widths for the $S$-wave states ($\eta_c, \eta_b$) and the $P$-wave states 
($\chi_{cJ}, \chi_{bJ}$ for $J=0,1,2$) radiative Dalitz decays into a lepton pair 
($e^{+}e^{-}, \mu^{+}\mu^{-}$) with a photon at NLO QCD accuracy within the framework 
of NRQCD factorization. We first evaluate the lepton-pair invariant mass distribution 
and find that in the $S$-wave cases, the predominant contribution comes from the 
low-invariant-mass threshold region caused by the almost on-shell intermediate photon, 
while for the $P$-wave $\chi_{Q0}$ and $\chi_{Q2}$ states, an additional considerable 
contribution also arises from the soft-photon region. However, for the $J=1$ 
$\chi_{Q1}$ states, only the soft-photon region contributes predominately due to the 
Landau-Yang theorem. Furthermore, comparing the electron and muon channels shows an 
interesting difference between lepton flavors. For the scalar ($J=0$) and tensor 
($J=2$) states, the muon modes are much more sensitive to the photon energy cut than 
the electron modes. This happens because the heavy muon mass ($m_\mu \gg m_e$) modifies 
the phase space near the endpoint and regularizes the collinear singularities. 
Remarkably, this lepton-flavor difference completely disappears for the axial-vector 
($J=1$) states. The $\chi_{c1}$ and $\chi_{b1}$ configurations yield completely 
identical decay widths in both electron and muon channels across all choices of 
$E_{\gamma}^{\rm cut}$. These spectacular properties lead to very rich 
$\gamma^{\ast}\to l^{+}l^{-}$ form factors that can be probed by experimental 
measurements.

To compare with the practical experimental measurements of the total widths of the Dalitz decay and to examine the predictive power of perturbative calculations, we further integrate over the whole phase space imposing three different cuts on 
the photon energy, which are $E_{\gamma}^{\rm cut} = 100$, $300$, and $500~\text{MeV}$. 
We find that for the $S$-wave states, the theoretical predictions almost do not change 
under these cut conditions. However, the total decay widths of the $P$-wave states show 
a clear sensitivity hierarchy ordered as $\chi_{Q1} > \chi_{Q2} > \chi_{Q0}$. 
Specifically, in the electron channel, the NLO decay widths drop significantly by $66\%~(50\%)$ for the $\chi_{c1}~(\chi_{b1})$ states. In contrast, the tensor $\chi_{Q2}$ states show a moderate decrease of $21\%~(18\%)$. Meanwhile, the scalar $\chi_{Q0}$ states are the most stable against the cut, with a small reduction of only $5\%$--$6\%$. For the $J=0$ and $J=2$ decays into the muon channels, the dependencies become even stronger. When $E_{\gamma}^{\rm cut}$ increases from $100~\text{MeV}$ to $500~\text{MeV}$, the NLO widths drop by $17\%~(11\%)$ for the $\chi_{c0}~(\chi_{b0})$ states and by $39\%~(29\%)$ for the $\chi_{c2}~(\chi_{b2})$ states. The results reveal that the NLO QCD radiative corrections introduce a major qualitative 
split among the multiplets, demonstrating that no simple, global $K$-factor can be 
applied uniformly to $P$-wave quarkonium Dalitz decays. 

Fortunately, implementing the same photon energy cut in both theory and experimental analyses allows for a direct, critical validation of our framework. This study opens clear avenues for upcoming high-precision experimental tests: we look forward to dedicated searches by the BESIII experiment for the $\chi_{cJ} \to e^+e^-\gamma$ and $\chi_{cJ} \to \mu^+\mu^-\gamma$ signals, as well as future measurements at Belle II, so that our theoretical predictions and the underlying infrared structure of NRQCD can be critically tested in the near future.

\hspace{2cm}

\noindent {\bf Acknowledgments:} We thank Xiao-Rui Lyu for his helpful suggestions. The work of Z.-G. H. is supported by the Fundamental Research Funds for the Central Universities through Grant No. buctrc202432; The work of X.-D. H. is supported by the National Natural Science Foundation of China under Grants No. 12505097, and the Chongqing Natural Science Foundation under Grant No. CSTB2025NSCQ-GPX1018.

\hspace{2cm}

\bibliographystyle{JHEP}
\bibliography{nrqcd}

@article{Bodwin:1994jh,
    author = "Bodwin, Geoffrey T. and Braaten, Eric and Lepage, G. Peter",
    title = "{Rigorous QCD analysis of inclusive annihilation and production of heavy quarkonium}",
    eprint = "hep-ph/9407339",
    archivePrefix = "arXiv",
    reportNumber = "ANL-HEP-PR-94-24, FERMILAB-PUB-94-073-T, NUHEP-TH-94-5",
    doi = "10.1103/PhysRevD.55.5853",
    journal = "Phys. Rev. D",
    volume = "51",
    pages = "1125--1171",
    year = "1995",
    note = "[Erratum: Phys.Rev.D 55, 5853 (1997)]"
}

@article{Brambilla:2010cs,
    author = "Brambilla, N. and others",
    title = "{Heavy Quarkonium: Progress, Puzzles, and Opportunities}",
    eprint = "1010.5827",
    archivePrefix = "arXiv",
    primaryClass = "hep-ph",
    reportNumber = "SLAC-R-996, TUM-EFT-11-10, CLNS-10-2066, ANL-HEP-PR-10-44, ALBERTA-THY-11-10, CP3-10-37, FZJ-IKP-TH-2010-24, INT-PUB-10-059, JLAB-THY-11-1308, FERMILAB-PUB-10-737-T",
    doi = "10.1140/epjc/s10052-010-1534-9",
    journal = "Eur. Phys. J. C",
    volume = "71",
    pages = "1534",
    year = "2011"
}

@article{BESIII:2014uzr,
    author = "Ablikim, M. and others",
    collaboration = "BESIII",
    title = "{Evidence for $e^+e^-\to\gamma\chi_{c1, 2}$ at center-of-mass energies from 4.009 to 4.360 GeV}",
    eprint = "1411.6336",
    archivePrefix = "arXiv",
    primaryClass = "hep-ex",
    doi = "10.1088/1674-1137/39/4/041001",
    journal = "Chin. Phys. C",
    volume = "39",
    pages = "041001",
    year = "2015"
}

@article{BESIII:2021yal,
    author = "Ablikim, M. and others",
    collaboration = "BESIII",
    title = "{Measurement of $e^+e^-\to\gamma \chi_{c0,c1,c2}$ cross sections at center-of-mass energies between 3.77 and 4.60~GeV}",
    eprint = "2107.03604",
    archivePrefix = "arXiv",
    primaryClass = "hep-ex",
    doi = "10.1103/PhysRevD.104.092001",
    journal = "Phys. Rev. D",
    volume = "104",
    pages = "092001",
    year = "2021"
}

@article{Belle:2018jqa,
    author = "Jia, S. and others",
    collaboration = "Belle",
    title = "{Observation of $e^+e^- \to \gamma \chi_{c1}$ and search for $e^+e^- \to \gamma \chi_{c0}, \gamma \chi_{c2},$ and $\gamma\eta_c$ at $\sqrt{s}$ near 10.6 GeV at Belle}",
    eprint = "1810.10291",
    archivePrefix = "arXiv",
    primaryClass = "hep-ex",
    reportNumber = "Belle Preprint {\#} 2018-22, KEK Preprint {\#} 2018-55",
    doi = "10.1103/PhysRevD.98.092015",
    journal = "Phys. Rev. D",
    volume = "98",
    pages = "092015",
    year = "2018"
}

@article{Li:2009ki,
    author = "Li, Dan and He, Zhi-Guo and Chao, Kuang-Ta",
    title = "{Search for C= charmonium and bottomonium states in $e^+ e^- \to \gamma + X$ at B factories}",
    eprint = "0910.4155",
    archivePrefix = "arXiv",
    primaryClass = "hep-ph",
    doi = "10.1103/PhysRevD.80.114014",
    journal = "Phys. Rev. D",
    volume = "80",
    pages = "114014",
    year = "2009"
}

@article{Li:2013nna,
    author = "Li, Yi-Jie and Xu, Guang-Zhi and Liu, Kui-Yong and Zhang, Yu-Jie",
    title = "{Search for $C=+$ charmonium and XYZ states in $e^+e^-\to \gamma+ H$ at BESIII}",
    eprint = "1310.0374",
    archivePrefix = "arXiv",
    primaryClass = "hep-ph",
    doi = "10.1007/JHEP01(2014)022",
    journal = "JHEP",
    volume = "01",
    pages = "022",
    year = "2014"
}

@article{Chao:2013cca,
    author = "Chao, Kuang-Ta and He, Zhi-Guo and Li, Dan and Meng, Ce",
    title = "{Search for $C=+$ charmonium states in $e^+e^-\to \gamma+~X$ at BEPCII/BESIII}",
    eprint = "1310.8597",
    archivePrefix = "arXiv",
    primaryClass = "hep-ph",
    reportNumber = "DESY-13-198",
    month = "10",
    year = "2013"
}

@article{Yuan:2015kya,
    author = "Yuan, Chang-Zheng",
    collaboration = "BESIII",
    title = "{Study of the XYZ states at the BESIII}",
    eprint = "1509.06850",
    archivePrefix = "arXiv",
    primaryClass = "hep-ex",
    doi = "10.1007/s11467-015-0484-y",
    journal = "Front. Phys. (Beijing)",
    volume = "10",
    pages = "101401",
    year = "2015"
}

@article{BESIII:2013fnz,
    author = "Ablikim, M. and others",
    collaboration = "BESIII",
    title = "{Observation of $e^+ e^- \to \gamma X(3872)$ at BESIII}",
    eprint = "1310.4101",
    archivePrefix = "arXiv",
    primaryClass = "hep-ex",
    doi = "10.1103/PhysRevLett.112.092001",
    journal = "Phys. Rev. Lett.",
    volume = "112",
    pages = "092001",
    year = "2014"
}

@article{Shifman:1980dk,
    author = "Shifman, Mikhail A. and Vysotsky, Michael I.",
    title = "{FORM-FACTORS OF HEAVY MESONS IN QCD}",
    reportNumber = "ITEP-147-1980",
    doi = "10.1016/0550-3213(81)90023-7",
    journal = "Nucl. Phys. B",
    volume = "186",
    pages = "475--518",
    year = "1981"
}

@article{Chung:2008km,
    author = "Chung, Hee Sok and Lee, Jungil and Yu, Chaehyun",
    title = "{Exclusive heavy quarkonium + gamma production from e+ e- annihilation into a virtual photon}",
    eprint = "0808.1625",
    archivePrefix = "arXiv",
    primaryClass = "hep-ph",
    doi = "10.1103/PhysRevD.78.074022",
    journal = "Phys. Rev. D",
    volume = "78",
    pages = "074022",
    year = "2008"
}

@article{Sang:2009jc,
    author = "Sang, Wen-Long and Chen, Yu-Qi",
    title = "{Higher Order Corrections to the Cross Section of e+e- ---{\ensuremath{>}} Quarkonium + gamma}",
    eprint = "0910.4071",
    archivePrefix = "arXiv",
    primaryClass = "hep-ph",
    doi = "10.1103/PhysRevD.81.034028",
    journal = "Phys. Rev. D",
    volume = "81",
    pages = "034028",
    year = "2010"
}

@article{Fan:2012dy,
    author = "Fan, Ying and Lee, Jungil and Yu, Chaehyun",
    title = "{Resummation of relativistic corrections to exclusive productions of charmonia in $e^+e^-$ collisions}",
    eprint = "1211.4111",
    archivePrefix = "arXiv",
    primaryClass = "hep-ph",
    reportNumber = "KIAS-PREPRINT-P12062, SLAC-PUB-15275",
    doi = "10.1103/PhysRevD.87.094032",
    journal = "Phys. Rev. D",
    volume = "87",
    pages = "094032",
    year = "2013"
}

@article{Xu:2014zra,
    author = "Xu, Guang-Zhi and Li, Yi-Jie and Liu, Kui-Yong and Zhang, Yu-Jie",
    title = "{$\alpha_sv^2$ corrections to $\eta_c$ and $\chi_{cJ}$ production recoiled with a photon at $e^+e^-$ colliders}",
    eprint = "1407.3783",
    archivePrefix = "arXiv",
    primaryClass = "hep-ph",
    doi = "10.1007/JHEP10(2014)071",
    journal = "JHEP",
    volume = "10",
    pages = "071",
    year = "2014"
}

@article{Brambilla:2017kgw,
    author = "Brambilla, Nora and Chen, Wen and Jia, Yu and Shtabovenko, Vladyslav and Vairo, Antonio",
    title = "{Relativistic corrections to exclusive $\chi_{cJ} + \gamma$ production from $e^+ e^-$ annihilation}",
    eprint = "1712.06165",
    archivePrefix = "arXiv",
    primaryClass = "hep-ph",
    reportNumber = "TUM-EFT-68-15, TUM-EFT 68/15",
    doi = "10.1103/PhysRevD.97.096001",
    journal = "Phys. Rev. D",
    volume = "97",
    pages = "096001",
    year = "2018",
    note = "[Erratum: Phys.Rev.D 101, 039903 (2020)]"
}

@article{Chen:2017pyi,
    author = "Chen, Long-Bin and Liang, Yi and Qiao, Cong-Feng",
    title = "{NNLO QCD corrections to $\gamma + \eta_c(\eta_b)$ exclusive production in electron-positron collision}",
    eprint = "1710.07865",
    archivePrefix = "arXiv",
    primaryClass = "hep-ph",
    doi = "10.1007/JHEP01(2018)091",
    journal = "JHEP",
    volume = "01",
    pages = "091",
    year = "2018"
}

@article{Yu:2020tri,
    author = "Yu, Huai-Min and Sang, Wen-Long and Huang, Xu-Dong and Zeng, Jun and Wu, Xing-Gang and Brodsky, Stanley J.",
    title = "{Scale-fixed predictions for $\gamma + \eta_c$ production in electron-positron collisions at NNLO in perturbative QCD}",
    eprint = "2007.14553",
    archivePrefix = "arXiv",
    primaryClass = "hep-ph",
    doi = "10.1007/JHEP01(2021)131",
    journal = "JHEP",
    volume = "01",
    pages = "131",
    year = "2021"
}

@article{Sang:2020fql,
    author = "Sang, Wen-Long and Feng, Feng and Jia, Yu",
    title = "{Next-to-next-to-leading-order radiative corrections to $e^+e^-\to\chi_{cJ}+\gamma$ at B factory}",
    eprint = "2008.04898",
    archivePrefix = "arXiv",
    primaryClass = "hep-ph",
    doi = "10.1007/JHEP10(2020)098",
    journal = "JHEP",
    volume = "10",
    pages = "098",
    year = "2020"
}

@article{Li:2025pbt,
    author = "Li, Cong and Sang, Wen-Long and Zhang, Hong-Fei",
    title = "{The next-to-next-to-leading-order QCD corrections to $e^+e^- \to \eta_c/\chi_{cJ}+\gamma$ at B factories}",
    eprint = "2512.04758",
    archivePrefix = "arXiv",
    primaryClass = "hep-ph",
    month = "12",
    year = "2025"
}

@article{Jia:2008ep,
    author = "Jia, Yu and Yang, Deshan",
    title = "{Refactorizing NRQCD short-distance coefficients in exclusive quarkonium production}",
    eprint = "0812.1965",
    archivePrefix = "arXiv",
    primaryClass = "hep-ph",
    doi = "10.1016/j.nuclphysb.2009.01.025",
    journal = "Nucl. Phys. B",
    volume = "814",
    pages = "217--230",
    year = "2009"
}

@article{Chung:2019ota,
    author = "Chung, Hee Sok and Ee, June-Haak and Kang, Daekyoung and Kim, U-Rae and Lee, Jungil and Wang, Xiang-Peng",
    title = "{Pseudoscalar Quarkonium+gamma Production at NLL+NLO accuracy}",
    eprint = "1906.03275",
    archivePrefix = "arXiv",
    primaryClass = "hep-ph",
    doi = "10.1007/JHEP10(2019)162",
    journal = "JHEP",
    volume = "10",
    pages = "162",
    year = "2019"
}

@article{BaBar:2010siw,
    author = "Lees, J. P. and others",
    collaboration = "BaBar",
    title = "{Measurement of the $\gamma \gamma* --> \eta_c$ transition form factor}",
    eprint = "1002.3000",
    archivePrefix = "arXiv",
    primaryClass = "hep-ex",
    reportNumber = "BABAR-PUB-09-034, SLAC-PUB-13953",
    doi = "10.1103/PhysRevD.81.052010",
    journal = "Phys. Rev. D",
    volume = "81",
    pages = "052010",
    year = "2010"
}

@article{Feng:2015uha,
    author = "Feng, Feng and Jia, Yu and Sang, Wen-Long",
    title = "{Can Nonrelativistic QCD Explain the $\gamma\gamma^* \to \eta_c$  Transition Form Factor Data?}",
    eprint = "1505.02665",
    archivePrefix = "arXiv",
    primaryClass = "hep-ph",
    doi = "10.1103/PhysRevLett.115.222001",
    journal = "Phys. Rev. Lett.",
    volume = "115",
    pages = "222001",
    year = "2015"
}

@article{Wang:2018lry,
    author = "Wang, Sheng-Quan and Wu, Xing-Gang and Sang, Wen-Long and Brodsky, Stanley J.",
    title = "{Solution to the $\gamma\gamma^*\rightarrow\eta_c$ puzzle using the principle of maximum conformality}",
    eprint = "1804.06106",
    archivePrefix = "arXiv",
    primaryClass = "hep-ph",
    reportNumber = "SLAC-PUB-17247",
    doi = "10.1103/PhysRevD.97.094034",
    journal = "Phys. Rev. D",
    volume = "97",
    pages = "094034",
    year = "2018"
}

@article{Babiarz:2025agk,
    author = {Babiarz, Izabela and Flett, Chris A. and Ozcelik, Melih A. and Sch{\"a}fer, Wolfgang and Szczurek, Antoni},
    title = "{Transition form-factor for $\eta _Q$ at NNLO in the strong coupling $\alpha _s$ and with all-order $v^2$ resummation}",
    eprint = "2509.15310",
    archivePrefix = "arXiv",
    primaryClass = "hep-ph",
    doi = "10.1140/epjc/s10052-025-15226-2",
    journal = "Eur. Phys. J. C",
    volume = "85",
    pages = "1474",
    year = "2025"
}

@article{Belle:2020ndp,
    author = "Teramoto, Y. and others",
    collaboration = "Belle",
    title = "{Evidence for $X(3872)\rightarrow J/\psi \pi^+\pi^-$ Produced in Single-Tag Two-Photon Interactions}",
    eprint = "2007.05696",
    archivePrefix = "arXiv",
    primaryClass = "hep-ex",
    reportNumber = "Belle Preprint 2020-08, KEK Preprint 2020-07",
    doi = "10.1103/PhysRevLett.126.122001",
    journal = "Phys. Rev. Lett.",
    volume = "126",
    pages = "122001",
    year = "2021"
}

@article{Babiarz:2023ebe,
    author = {Babiarz, Izabela and Pasechnik, Roman and Sch{\"a}fer, Wolfgang and Szczurek, Antoni},
    title = "{Probing the structure of {\ensuremath{\chi}}c1(3872) with photon transition form factors}",
    eprint = "2303.09175",
    archivePrefix = "arXiv",
    primaryClass = "hep-ph",
    doi = "10.1103/PhysRevD.107.L071503",
    journal = "Phys. Rev. D",
    volume = "107",
    pages = "L071503",
    year = "2023"
}

@article{DiSalvo:2000ec,
    author = "Di Salvo, Elvio and Rekalo, Michail P. and Tomasi-Gustafsson, Egle",
    title = "{The eta(c) gamma gamma* transition form-factor in the decay eta(c) ---{\ensuremath{>}} gamma lepton+ lepton- and in the crossed channels gamma e- ---{\ensuremath{>}} eta(c) e- and e+ e- ---{\ensuremath{>}} eta(c) gamma}",
    eprint = "hep-ph/0004112",
    archivePrefix = "arXiv",
    reportNumber = "GEF-TH-3-2000A",
    doi = "10.1007/s100520050022",
    journal = "Eur. Phys. J. C",
    volume = "16",
    pages = "295--302",
    year = "2000"
}

@article{Jia:2009ip,
    author = "Jia, Yu and Sang, Wen-Long",
    title = "{Observation prospects of leptonic and Dalitz decays of pseudoscalar quarkonia}",
    eprint = "0906.4782",
    archivePrefix = "arXiv",
    primaryClass = "hep-ph",
    doi = "10.1088/1126-6708/2009/10/090",
    journal = "JHEP",
    volume = "10",
    pages = "090",
    year = "2009"
}

@article{Feng:2017hlu,
    author = "Feng, Feng and Jia, Yu and Sang, Wen-Long",
    title = "{Next-to-Next-to-Leading-Order QCD Corrections to the Hadronic width of Pseudoscalar Quarkonium}",
    eprint = "1707.05758",
    archivePrefix = "arXiv",
    primaryClass = "hep-ph",
    doi = "10.1103/PhysRevLett.119.252001",
    journal = "Phys. Rev. Lett.",
    volume = "119",
    pages = "252001",
    year = "2017"
}

@article{ParticleDataGroup:2024cfk,
    author = "Navas, S. and others",
    collaboration = "Particle Data Group",
    title = "{Review of particle physics}",
    doi = "10.1103/PhysRevD.110.030001",
    journal = "Phys. Rev. D",
    volume = "110",
    pages = "030001",
    year = "2024"
}

@article{Sang:2015uxg,
    author = "Sang, Wen-Long and Feng, Feng and Jia, Yu and Liang, Shuang-Ran",
    title = "{Next-to-next-to-leading-order QCD corrections to $\chi_{c0,2}\rightarrow \gamma\gamma$}",
    eprint = "1511.06288",
    archivePrefix = "arXiv",
    primaryClass = "hep-ph",
    doi = "10.1103/PhysRevD.94.111501",
    journal = "Phys. Rev. D",
    volume = "94",
    pages = "111501",
    year = "2016"
}

@article{BESIII:2026pff,
    author = "Ablikim, Medina and others",
    collaboration = "BESIII",
    title = "{First Measurement of the Absolute Branching Fraction of $eta_c \to \gamma\gamma$}",
    eprint = "2601.11236",
    archivePrefix = "arXiv",
    primaryClass = "hep-ex",
    month = "1",
    year = "2026"
}

@article{Bodwin:2007zf,
    author = "Bodwin, Geoffrey T. and Braaten, Eric and Kang, Daekyoung and Lee, Jungil",
    title = "{Inclusive charm production in chi(b) decays}",
    eprint = "0704.2599",
    archivePrefix = "arXiv",
    primaryClass = "hep-ph",
    reportNumber = "ANL-HEP-PR-07-20",
    doi = "10.1103/PhysRevD.76.054001",
    journal = "Phys. Rev. D",
    volume = "76",
    pages = "054001",
    year = "2007"
}

@article{Yang:2012gk,
    author = "Yang, Deshan and Zhao, Shuai",
    title = "{$\chi_{QJ} \to l^+l^-$ within and beyond the Standard Model}",
    eprint = "1203.3389",
    archivePrefix = "arXiv",
    primaryClass = "hep-ph",
    doi = "10.1140/epjc/s10052-012-1996-z",
    journal = "Eur. Phys. J. C",
    volume = "72",
    pages = "1996",
    year = "2012"
}

@article{Kivel:2015iea,
    author = "Kivel, N. and Vanderhaeghen, M.",
    title = "{$\chi_{cJ}\rightarrow e^{+}e^{-}$ decays revisited}",
    eprint = "1509.07375",
    archivePrefix = "arXiv",
    primaryClass = "hep-ph",
    doi = "10.1007/JHEP02(2016)032",
    journal = "JHEP",
    volume = "02",
    pages = "032",
    year = "2016"
}

@article{Jia:2024dzm,
    author = "Jia, Yu and Pan, Jichen",
    title = "{A tale of Bethe logarithms: leptonic widths of $\chi_{cJ}$ and Lamb shift}",
    eprint = "2411.18560",
    archivePrefix = "arXiv",
    primaryClass = "hep-ph",
    month = "11",
    year = "2024"
}

@article{Li:2009ad,
    author = "Li, Bai-Qing and Meng, Ce and Chao, Kuang-Ta",
    title = "{Coupled-Channel and Screening Effects in Charmonium Spectrum}",
    eprint = "0904.4068",
    archivePrefix = "arXiv",
    primaryClass = "hep-ph",
    doi = "10.1103/PhysRevD.80.014012",
    journal = "Phys. Rev. D",
    volume = "80",
    pages = "014012",
    year = "2009"
}

@article{Wang:2015rcz,
    author = "Wang, Wei and Zhao, Qiang",
    title = "{Decipher the short-distance component of $X(3872)$ in $B_c$ decays}",
    eprint = "1512.03123",
    archivePrefix = "arXiv",
    primaryClass = "hep-ph",
    doi = "10.1016/j.physletb.2016.02.012",
    journal = "Phys. Lett. B",
    volume = "755",
    pages = "261--264",
    year = "2016"
}

@article{Tan:2019qwe,
    author = "Tan, Yue and Ping, Jialun",
    title = "{X(3872) in an unquenched quark model}",
    eprint = "1906.09690",
    archivePrefix = "arXiv",
    primaryClass = "hep-ph",
    doi = "10.1103/PhysRevD.100.034022",
    journal = "Phys. Rev. D",
    volume = "100",
    pages = "034022",
    year = "2019"
}

@article{Man:2024mvl,
    author = "Man, Zi-Long and Shu, Cheng-Rui and Liu, Yan-Rui and Chen, Hong",
    title = "{Charmonium states in a coupled-channel model}",
    eprint = "2402.02765",
    archivePrefix = "arXiv",
    primaryClass = "hep-ph",
    doi = "10.1140/epjc/s10052-024-13132-7",
    journal = "Eur. Phys. J. C",
    volume = "84",
    pages = "810",
    year = "2024"
}

@article{Liu:2022chg,
    author = "Liu, Xiao and Ma, Yan-Qing",
    title = "{AMFlow: A Mathematica package for Feynman integrals computation via auxiliary mass flow}",
    eprint = "2201.11669",
    archivePrefix = "arXiv",
    primaryClass = "hep-ph",
    doi = "10.1016/j.cpc.2022.108565",
    journal = "Comput. Phys. Commun.",
    volume = "283",
    pages = "108565",
    year = "2023"
}

@article{Liu:2017jxz,
    author = "Liu, Xiao and Ma, Yan-Qing and Wang, Chen-Yu",
    title = "{A Systematic and Efficient Method to Compute Multi-loop Master Integrals}",
    eprint = "1711.09572",
    archivePrefix = "arXiv",
    primaryClass = "hep-ph",
    doi = "10.1016/j.physletb.2018.02.026",
    journal = "Phys. Lett. B",
    volume = "779",
    pages = "353--357",
    year = "2018"
}

@article{Liu:2020kpc,
    author = "Liu, Xiao and Ma, Yan-Qing and Tao, Wei and Zhang, Peng",
    title = "{Calculation of Feynman loop integration and phase-space integration via auxiliary mass flow}",
    eprint = "2009.07987",
    archivePrefix = "arXiv",
    primaryClass = "hep-ph",
    doi = "10.1088/1674-1137/abc538",
    journal = "Chin. Phys. C",
    volume = "45",
    pages = "013115",
    year = "2021"
}

@article{Liu:2021wks,
    author = "Liu, Xiao and Ma, Yan-Qing",
    title = "{Multiloop corrections for collider processes using auxiliary mass flow}",
    eprint = "2107.01864",
    archivePrefix = "arXiv",
    primaryClass = "hep-ph",
    doi = "10.1103/PhysRevD.105.L051503",
    journal = "Phys. Rev. D",
    volume = "105",
    pages = "L051503",
    year = "2022"
}

@article{Hahn:2000kx,
    author = "Hahn, Thomas",
    title = "{Generating Feynman diagrams and amplitudes with FeynArts 3}",
    eprint = "hep-ph/0012260",
    archivePrefix = "arXiv",
    reportNumber = "KA-TP-23-2000",
    doi = "10.1016/S0010-4655(01)00290-9",
    journal = "Comput. Phys. Commun.",
    volume = "140",
    pages = "418--431",
    year = "2001"
}

@article{Klappert:2020nbg,
    author = {Klappert, Jonas and Lange, Fabian and Maierh\"ofer, Philipp and Usovitsch, Johann},
    title = "{Integral reduction with Kira 2.0 and finite field methods}",
    eprint = "2008.06494",
    archivePrefix = "arXiv",
    primaryClass = "hep-ph",
    reportNumber = "TTK-20-24, P3H-20-041, FR-PHENO-2020-11, MITP/20-044",
    doi = "10.1016/j.cpc.2021.108024",
    journal = "Comput. Phys. Commun.",
    volume = "266",
    pages = "108024",
    year = "2021"
}

@article{Herren:2017osy,
    author = "Herren, Florian and Steinhauser, Matthias",
    title = "{Version 3 of RunDec and CRunDec}",
    eprint = "1703.03751",
    archivePrefix = "arXiv",
    primaryClass = "hep-ph",
    reportNumber = "TTP17-011",
    doi = "10.1016/j.cpc.2017.11.014",
    journal = "Comput. Phys. Commun.",
    volume = "224",
    pages = "333--345",
    year = "2018"
}

@article{Eichten:1995ch,
    author = "Eichten, Estia J. and Quigg, Chris",
    title = "{Quarkonium wave functions at the origin}",
    eprint = "hep-ph/9503356",
    archivePrefix = "arXiv",
    reportNumber = "FERMILAB-PUB-95-045-T, CLNS-95-1329",
    doi = "10.1103/PhysRevD.52.1726",
    journal = "Phys. Rev. D",
    volume = "52",
    pages = "1726--1728",
    year = "1995"
}

@article{Petrelli:1997ge,
    author = "Petrelli, Andrea and Cacciari, Matteo and Greco, Mario and Maltoni, Fabio and Mangano, Michelangelo L.",
    title = "{NLO production and decay of quarkonium}",
    eprint = "hep-ph/9707223",
    archivePrefix = "arXiv",
    reportNumber = "CERN-TH-97-142, DESY-97-090",
    doi = "10.1016/S0550-3213(97)00801-8",
    journal = "Nucl. Phys. B",
    volume = "514",
    pages = "245--309",
    year = "1998"
}

@article{Hahn:1998yk,
    author = "Hahn, T. and Perez-Victoria, M.",
    title = "{Automatized one loop calculations in four-dimensions and D-dimensions}",
    eprint = "hep-ph/9807565",
    archivePrefix = "arXiv",
    reportNumber = "UG-FT-87-98, KA-TP-7-1998",
    doi = "10.1016/S0010-4655(98)00173-8",
    journal = "Comput. Phys. Commun.",
    volume = "118",
    pages = "153--165",
    year = "1999"
}

@article{Lepage:1977sw,
    author = "Lepage, G. Peter",
    title = "{A New Algorithm for Adaptive Multidimensional Integration}",
    reportNumber = "SLAC-PUB-1839-REV, SLAC-PUB-1839",
    doi = "10.1016/0021-9991(78)90004-9",
    journal = "J. Comput. Phys.",
    volume = "27",
    pages = "192",
    year = "1978"
}

@article{Liu:2022mfb,
    author = "Liu, Zhi-Feng and Ma, Yan-Qing",
    title = "{Determining Feynman Integrals with Only Input from Linear Algebra}",
    eprint = "2201.11637",
    archivePrefix = "arXiv",
    primaryClass = "hep-ph",
    doi = "10.1103/PhysRevLett.129.222001",
    journal = "Phys. Rev. Lett.",
    volume = "129",
    pages = "222001",
    year = "2022"
}

\end{document}